\documentclass[a4paper,fleqn,usenatbib]{mnras}

\usepackage{newtxtext,newtxmath}
\usepackage[T1]{fontenc}
\usepackage{ae,aecompl}

\usepackage{graphicx}	% Including figure files
\usepackage{amsmath}	% Advanced maths commands
\usepackage{amssymb}	% Extra maths symbols
\usepackage{bm}
\usepackage{tabularx} 
\title[Production of intense episodic Alfv\'en pulses
]
{Production of intense episodic Alfv\'en pulses:
GRMHD simulation of black hole accretion disks}

\author[A. Mizuta et al.]{
Akira Mizuta,$^{1}$\thanks{E-mail: akira.mizuta@riken.jp}
Toshikazu Ebisuzaki$^{1}$,
Toshiki Tajima$^{2}$,
and Shigehiro Nagataki$^{3,4}$
\\
$^{1}$Computational Astrophysics Laboratory, RIKEN, 2-1
Hirosawa, Wako, Saitama,
351-0198, Japan\\
$^{2}$Department of Physics and Astronomy, University of California at Irvine, Irvine, CA 92697, United States\\
$^{3}$Astrophysical Big Bang Laboratory, RIKEN Cluster for Pioneering Research, RIKEN, 2-1
Hirosawa, Wako, Saitama,
351-0198, Japan\\
$^{4}$RIKEN Interdisciplinary Theoretical \& Mathematical Science Program (iTHEMS), 2-1 Hirosawa, Wako, Saitama, 351-0198 Japan
}

\date{Accepted XXX. Received YYY; in original form ZZZ}

\pubyear{2018}

\begin{document}
\label{firstpage}
\pagerange{\pageref{firstpage}--\pageref{lastpage}}
\maketitle

\begin{abstract}
The episodic dynamics of the magnetic eruption of a spinning black hole (BH)
accretion disks and its associated intense
shapeup of their jets is studied via
three-dimensional general-relativistic magnetohydrodynamics (GRMHD).
The embedded magnetic fields in the disk
get amplified by the magnetorotational instability (MRI)
so large as to cause an eruption of magnetic field (recconection) 
and large chunks of matter episodically accrete
toward the roots of the jets upon such an event.
We also find that the eruption events produce intensive 
Alfv\'en pulses, which propagate through the jets.
 After the eruption, the disk backs to the weakly magnetic states.
Such disk activities  cause 
short time variabilities in mass accretion rate at the event horizon
as well as electromagnetic luminosity inside the jet.
Since the dimensionless strength parameter $a_0=eE/m_e \omega c$
of these Alfv\'en wave pulses is
extremely high
for a substantial fraction of Eddington accretion rate accretion flow
onto a supermassive black hole,
the Alfv\'en shocks turn into ultrarelativistic $(a_0\gg 1)$
bow wake acceleration,
manifesting  into
the ultra-high energy cosmic rays
and electrons which finally emit gamma-rays.
Since our GRMHD model has universality in its spatial and
temporal scales,
it is applicable to a wide range of astrophysical objects
ranging from those of AGN 
(which is the primary target of this research),
to micro-quasars.
Properties such as
time variabilities of blazar gamma-ray flares 
and spectrum observed by {\it Fermi} Gamma-ray Observatory are well explained
by linear acceleration of electrons by the bow wake.
\end{abstract}


%%%%%%%%%%%%%%%%%%%%%%%%%%%%%%%%%%%%%%%%%%%%%%%%%%

%%%%%%%%%%%%%%%%% BODY OF PAPER %%%%%%%%%%%%%%%%%%

\section{Introduction}

\label{sec:intro}
Active galactic Nuclei (AGNs) are high energy astronomical objects,
so that
they emit non-thermal radiation in any frequency ranges of radiation,
in other words, radio, infrared,
visible lights, ultraviolet, X-rays, and gamma-rays
\citep{Begelman84,Hughes91,Burgarella93,Tsiganos96,Ferrari98}.
These central engines are believed to be accreting supermassive black
holes,
with relativistic jet whose bulk Lorentz factor is $\sim 10$
\citep{Biretta99,Asada14,Boccardi16}.
Their jets show strong time variabilities in the timescales
from days to years \citep{Fossati98,Abdo09a,Abdo10a,Abdo10b,Abdo10c,Abdo11,Ackermann10,Chen13,Edelson13}.
In the extreme cases, blazars show bursts of hours \citep{Ackermann16,Britto16}.
These radiations are believed to be emitted by a bunch of electrons
with strongly relativistic motions.

The system of AGN jets is also believed as a cosmic ray accelerator
\citep{Dermer09}.
Although it has a potential to accelerate
highest energy to the energy of $\sim 10^{20}$ eV,
it is still not well understood
what is physical mechanism for the particle acceleration
and
where the acceleration site is.
Many studies 
have been done for diffusive shock acceleration model
based on conventional Fermi acceleration mechanism \citep{Fermi54}.
In the Fermi acceleration, charged particles interact
with magnetized clouds, and are
vented random directions by their magnetic field.
Since the head-on collisions, which the particles gain the energy,
are more frequent than rear-end collisions, which they loose their
energies,
the particles statistically gain the energy step by step and
eventually obtain very high energy to be cosmic ray particles.
However, this Fermi acceleration mechanism is
difficult to explain highest energy particles $\sim 10^{20}$ eV, 
because of 1) the large number of scatterings necessary
to reach highest energies,
2) energy losses through the synchrotron emission
at the bending associated with scatterings,
and 3) difficulty in the escape of particles
which are initially magnetically confined
in the acceleration domain (e.g., \citet{Kotera11}).

On the other hand, \citet{Tajima79} proposed that particles
can be accelerated by the wakefield induced by an intense laser pulse,
see review by \citet{Tajima17}.
This long lasting energy elevated state of wakefield may be
regarded a Higgs state \citep{Higgs64} of plasma.
In particular, ponderomotive force
which is proportional to gradient of ${ \bm{E}^2}$ works
to accelerate the charged particles, effectively,
where $\bm{E}$ is electric field of the electromagnetic wave.
The acceleration towards relativistic regime
by the ponderomotive force
is confirmed by recent experiments by ultra intense lasers
for electrons \citep{Leemans06,Nakamura07}.
Positron acceleration driven by wakefields by electrons
is also reported by \citet{Corde15}.

\citet{Takahashi00,Chen02,Chang09} applied 
this mechanism to magnetowave induced plasma wakefield acceleration
for ultra high energy cosmic rays.
Recently, \citet{Ebisuzaki14,Ebisuzaki14b} applied
this wakefield acceleration theory
to the relativistic jets launched
from an accreting black hole.
In such astrophysical context going far beyond the laboratory scales
of wakefields (see, for example, \citet{Tajima17}),
the relativistic factors that characterize the dynamics
$a_0=eE/ m_e \omega c$ becomes
far greater than unity.
This has not been achieved in the laboratory yet,
though simulations began to peek into it.
In this regime \citep{Ebisuzaki14,Ebisuzaki14b}
the ponderomotive acceleration has advantages over the Fermi mechanism. 
In close scrutiny, the wakefields are composed with
two parts the frontal bow part and the following stern (wake)
part.
Here we may call simply the bow wake and stern wake.
In the work of high $a_0$ simulation,
it was shown in \citet{Lau15} that the bow wake 
(it is driven directly by the ponderomotive force)
is dominant over the stern wake.
The advantages of this bow wake
acceleration over the Fermi mechanism are:
\begin{enumerate}
\item The ponderomotive field provides an extremely high accelerating field (including
the wakefield).
\item It does not require particle bending, which would cause strong synchrotron radiation
losses in extreme energies.
\item The accelerating fields and particles move in the collinear direction at the same
velocity, the speed of light, so that the acceleration has a built-in coherence called
``relativistic coherence'' \citep{Tajima2010}; in contrast, the Fermi acceleration mechanism, based
on multiple scatterings, is intrinsically incoherent and stochastic.
\item No escape problem \citep{Kotera11} exists. Particles can escape from the acceleration region
since the accelerating fields naturally decay out.
\end{enumerate}

They found that protons can be accelerated even above ZeV
$\sim 10^{22}$ eV in the bow wake of a burst of Alfv\'en waves emitted
by an accretion disk around a black hole with the mass of
$10^8 $ M$_{\odot}$.
\citet{Ebisuzaki14} used three major assumptions
based on the standard
$\alpha$-disk model \citep{Shakura73}.

\begin{itemize}
  \item {Assumption A}: the magnetic field energy ${\mathcal E}_B$
included in an Alfv\'en wave burst is assumed as:
\begin{equation}
{{\mathcal E}_B}=(B_D^2/4\pi)\pi(10R_s)^2Z_D=1.6\times 10^{48}(\dot{m}/0.1)(m/10^8)^2 \quad{\rm erg},
\label{eqn:EB}
\end{equation}
where $B_{D}$ is the magnetic field stored in the inner most regions of
the accretion disk, $R_{\rm s}=2R_{\rm g}$ is the Schwarzschild radius of the
black hole, $Z_{\rm D}$ is the thickness of the disk,
$R_{\rm g}$ is gravitational radius,
 $\dot{m}$ is
the
accretion rate normalized by the Eddington luminosity,
and $m$ is the mass of the black hole in the unit of solar mass.

\item{Assumption B}: they assumed that the angular frequency $\omega_{\rm A}$ of the
Alfv\'en wave corresponds to that excited by magnetorotational instability
(MRI \citep{Velikhov59,Chandrasekhar60,Balbus91,Matsumoto95}),
which takes place in a magnetized accretion disk, in other words:
\begin{equation}
\label{growthrate}
\omega_{\rm A}=2\pi {{c_{\rm A}}_D}/\lambda_{\rm A}\sim 2.6\times10^{-5}(m/10^8)^{-1}\quad {\rm Hz},
\end{equation}
where $\lambda_{\rm A}$ is the wavelength of the Alfv\'en wave,
and ${c_{\rm A}}$ is speed of Alfv\'en wave.
\citet{Ebisuzaki14} showed that the Alfv\'en shock gives rise to
electromagnetic wave pulse with $\omega=\omega_{\rm A}$
along the propagation in the jets through
the mode conversion, as the density and magnetic fields
in the jets decrease during the jet propagation.

\item{Assumption C}: the recurrence rate $\nu_{\rm A}$ of the Alfv\'en burst is evaluated as:
\begin{equation}
\label{recurrencerate}
\nu_{\rm A}=\eta {{c_{\rm A}}_D}/Z_{\rm D}\quad {\rm Hz},
\end{equation}
where $\eta$ is the episode-dependent parameter of the order of unity.
\end{itemize}

They found that the non-dimensional strength parameter 
$a_0=e E / m_e \omega c$ is as high as $10^{10}$
for the case of $\dot{m}=0.1$ and $m=10^8$,
where $e$ is electric charge, $E$ is the intensity of the electric
field,
$m_e$ is mass of electron, and $c$ is speed of light.
The ponderomotive force of this extremely relativistic waves co-linearly
accelerate to the jet particle up to the maximum energy:
\begin{equation}
W_{\rm max}=2.9\times 10^{22}q(\Gamma/20)
(\dot{m}/0.1)^{4/3}(m/10^8)^{2/3}\quad {\rm eV},
\end{equation}
where $q$ is the charge of the particle and $\Gamma$ is the bulk Lorentz
factor of the jet.
Recent one-dimensional particle in cell (PIC)
simulation shows maximum energy gain
via a ponderomotive force in the bow wake 
and the maximum energy is almost proportional to $a_0^2$
\citep{Lau15}.
Based on the above estimation, \citet{Ebisuzaki14}
concluded that the accreting supermassive black hole
is the ZeV ($10^{22}$ eV) linear accelerator. 

GRMHD simulations of accretion flows onto the black hole have been done
since early works by \citet{Koide99,Koide00}.
Some improvements in the numerical method for solving GRMHD equations
made it possible to follow the long-term dynamics of
magnetized accretion flows \citep{DeVilliers03,Gammie03}.
2D axis-symmetric and full 3D simulations have been done
to study properties of accretion disk, Blandford-Znajek efficiency,
jet and so on.

\citet{McKinney04} have studied outward going electromagnetic power through the event horizon,
i.e, 
Blandford-Znajek process by 2D axis-symmetric simulations.
\citet{McKinney06} studied long term magnetized jet propagation launched from the disk and black hole system.
The properties of the accretion flows, and outflows
are studied by a series of papers
\citep{DeVilliers03,Hirose04,DeVilliers05,Krolik05}
Long-term 3D GRMHD simulations have been done by \citet{Narayan12,McKinney12}
\citet{Beckwith08} pointed out that
the initial magnetic field topology strongly affects
the results, especially outflows, see also \citet{Narayan12,Penna13,Foucart16}.
Recently \citet{Ressler15,O'Riordan16,Chandra17,Shiokawa17} have tried to compare  the results by GRMHD simulations with observational results.

Major motivation of the present paper is to verify the assumptions of
\citet{Ebisuzaki14} (Equations (\ref{growthrate}) and
(\ref{recurrencerate})) by the 3D GRMHD
simulations of accretion disk around a supermassive black hole.
This paper is organized as follows.
We describe our physical models and numerical details in \S~2. 
The results are shown in \S~\ref{results}.
The application to the ultra high energy cosmic ray
acceleration, blazars, and gravitational waves
is discussed in \S~\ref{sec:discussion} and \S~5.

\section{GENERAL RELATIVISTIC MAGNETOHYDRODYNAMIC SIMULATION
METHOD}
\label{sec:basic_eq}
\subsection{Basic Equations}
\label{sec:basic_eq2}
We numerically solve general relativistic magnetohydrodynamic equations,
assuming a fixed metric around a black hole.
The unit in which $GM_{\rm BH}$ and $c$ are unity
is adopted, where $G$ is gravitational constant,
$M_{\rm BH}$ is mass of the central black hole.
The scales of length and time are
$R_{\rm g}=R_{\rm s}/2=GM_{\rm BH} c^{-2}$ and $GM_{\rm BH} c^{-3}$,
respectively.
The mass and energy is scale free.
The achieved mass accretion rate at the event horizon
is used to scale of the mass and energy, for example. 
The metric around a rotating black hole whose dimensionless
spin parameter is $a$
can be described by Boyer-Lindquist (BL)
coordinate or Kerr-Schild (KS) coordinate as follows.
\label{model}
The line element in BL coordinate is
\begin{eqnarray}
\label{BLmetric}
ds_{\rm BL}^2={g_{tt}}dt^2+g_{rr}dr^2+g_{\theta\theta}d\theta^2
+g_{\phi\phi}d\phi^2+2g_{t\phi}dtd\phi,
\end{eqnarray}
where $g_{tt}=-(1-2r/\Sigma)$, $g_{rr}=\Sigma/\Delta$,
$g_{\theta\theta}=\Sigma$, $g_{\phi\phi}=A\sin^2\theta/\Sigma$,
$g_{t\phi}=-2ra\sin^2\theta/\Sigma$,
$\Sigma=(r^2+a^2\cos^2 \theta)$,
$\Delta=r^2-2r+a^2$,
and $A=(r^2+a^2)^2-a^2\Delta\sin^2\theta$.
We follow standard notation used in \citet{Misner73},
i.e, metric tensor for Minkowski space is diag$(-1,1,1,1)$.

The line element in Kerr-Schild coordinate is
\begin{eqnarray}
\label{KSmetric}
ds_{\rm KS}^2={g_{tt}}dt^2+g_{rr}dr^2+g_{\theta\theta}d\theta^2
+g_{\phi\phi}d\phi^2+2g_{tr}dtdr \nonumber \\
+2g_{t\phi}dtd\phi+2g_{r\phi}drd\phi,
\end{eqnarray}
where $g_{tt}=-(1-2r/\Sigma)$, $g_{rr}=1+2r/\Sigma$,
$g_{\theta\theta}=\Sigma$, $g_{\phi\phi}=A\sin^2\theta/\Sigma$,
$g_{tr}=2r/\Sigma$,
$g_{t\phi}=-2ra\sin^2\theta/\Sigma$,
and $g_{r\phi}=-a\sin^2\theta(1+2r/\Sigma)$.

We also use so-called modified Kerr-Schild (mKS)
coordinate $(x_0,x_1,x_2,x_3)$
so that the numerical grids are fine near the event horizon
and the equator.
The transformation between
Kerr-Schild coordinate and modified Kerr-Schild coordinate
is described as
$t=x_0$, $r=\exp{x_1}$,
$\theta=\pi x_2 +{1\over 2}(1-h)\sin(2\pi x_2)$, and $\phi=x_3$,
where $h$ is a parameter which controls how the grids are
concentrated around the equator.
We have done three cases of resolution in the polar and azimuthal
angle grid.
Constant grid for polar angle, i.e., $h=1$
are used for two lower resolution cases.
The grid numbers are 
$N_1=124$, $N_2=124$, and $N_3=60$, 
and 
$N_1=124$, $N_2=252$, and $N_3=28$
which are uniformly spaced for both cases.
In another case we set $h=0.2$ so that
the polar grids concentrates around the equator.
The grid numbers are $N_1=124$, $N_2=252$, and $N_3=60$
which are uniformly spaced.
Since the resolution of polar grid with $h=0.2$ is
about 10 and 5 times better than
that with $h=1$ at the equator for the cases
with $N_2=124$ and $N_2=252$, respectively,
we can capture shorter wavelength 
and faster growing mode of MRI for poloidal direction
by the highest resolution case.
The highest resolution is comparable to those used in recent 3D GRMHD simulations 
\citep{McKinney12,Penna10}.
Computational domain covers from inside the event horizon
to $r=30000 R_{\rm g}$,
$[0.01\pi, 0.99\pi]$ in polar angle, and $[0, 2\pi]$ in azimuthal angle.
Contravariant vectors in Boyer-Lindquist coordinate
and Kerr-Schild coordinate are related with
$u^{t}_{KS}=u^t_{BL}+(2r/\Delta)u^{r}_{BL}$,
$u^{r}_{KS}=u^r_{BL}$,
$u^{\theta}_{KS}=u^{\theta}_{BL}$,
and $u^{\phi}_{KS}=(a/\Delta)u^r_{BL}+u^{\phi}_{BL}$.
Contravariant vectors in Kerr-Schild coordinate 
and modified Kerr-Schild coordinate are related with
$u^{t}_{KS}=u^0_{mKS}$,
$u^{r}_{KS}=ru^1_{mKS}$,
$u^{\theta}_{KS}=(\pi(1-(1-h)\cos(2\pi x_2)))u^2_{mKS}$,
and $u^{\phi}_{KS}=u^3_{mKS}$.

Mass and energy-momentum conservation laws are,
\begin{eqnarray}
\label{mass_conservation}
{1\over \sqrt{-g}}\partial_\mu (\sqrt{-g}\rho u^\mu)=0,\\
\label{enery_momentum_conservation}
\partial_\mu (\sqrt{-g}T^\mu _\nu)=\sqrt{-g}T^\kappa_\lambda
{\Gamma^\lambda} _{\nu\kappa},
\end{eqnarray}
where $T^{\mu \nu}$ is energy momentum tensor,
$\rho$ is rest mass density,
$u^\mu$ is fluid 4-velocity,
$g$ is determinant of metric tensor,
i.e., $g_{BL}=g_{KS}=-\Sigma^2\sin^2\theta$,
and $g_{mKS}=-\pi^2r^2(1-(1-h)\cos{2\pi x_2})^2\Sigma^2\sin^2\theta$,
and
$\Gamma ^k _{ij}$ is Christoffel symbol
which is defined as
$\Gamma ^k _{ij} =(1/2) g^{kl}
(\partial g_{jl}/\partial x^i
+\partial g_{li}/\partial x^j
-\partial g_{ij}/\partial x^l)$.
Energy momentum tensor which includes matter and electromagnetic parts
is defined as follows
\begin{eqnarray}
T^{\mu \nu}=T_{\rm MA}^{\mu \nu}+T_{\rm EM}^{\mu \nu},\\
T_{\rm MA}^{\mu \nu}=\rho h u^\mu u^\nu + p_{\rm th} g^{\mu \nu},\\
T_{\rm EM}^{\mu \nu}=F^{\mu\gamma}F^{\nu}_{\gamma}
-{1\over 4}g^{\mu\nu}F^{\alpha\beta}F_{\alpha\beta},
\end{eqnarray}
where $h(\equiv 1+U/\rho+p_{\rm th}/\rho)$ is specific enthalpy,
$U$ is thermal energy density,
$p_{\rm th}$ is thermal pressure, and
$F^{\mu\nu}$ is the Faraday tensor and a factor of $\sqrt{4\pi}$
is absorbed into the definition of $F^{\mu\nu}$.
The dual of the Faraday tensor is
\begin{eqnarray}
\label{Faraday_tensor}
^{*}F^{\mu\nu}={1\over 2}e^{\mu\nu\alpha\beta}F_{\alpha\beta},
\end{eqnarray}
where $e^{\mu\nu\alpha\beta}=\sqrt{-g}\epsilon^{\mu\nu\alpha\beta}$
and $\epsilon^{\mu\nu\alpha\beta}$ is the
completely antisymmetric Levi-Civita symbol
($\epsilon^{0123}=-\epsilon_{0123}=-1$).
The magnetic field observed by normal observer is
\begin{eqnarray}
{\mathcal B}^{\mu}=-^{*}F^{\mu\nu}n_{\nu}=\alpha~ {^{*}F^{\mu t}},
\end{eqnarray}
where $n_\mu=(-\alpha, 0,0,0)$ is the normal observer's four-velocity
and $\alpha \equiv\sqrt{-1/g^{tt}}$ is the lapse.
Note the time component of ${\mathcal B}$ is zero,
since ${\mathcal B}^t=\alpha~ ^*{F^{tt}=0}$.
Here we introduce another magnetic field which is used in \citet{Gammie03,Noble06,Nagataki09}
as 
\begin{eqnarray}
\label{mag}
B^{\mu}={^{*}F^{\mu t}}={{\mathcal B}^{\mu} \over \alpha}.
\end{eqnarray}
The time component of $B^\mu$ is also zero.
We also introduce four magnetic field $b^{\mu}$ 
which is measured by an observer at rest in the fluid,
\begin{eqnarray}
b^{\mu}=-{^{*}F^{\mu\nu}}u_{\nu}.
\end{eqnarray}
$B^{i}$ and $b^{\mu}$ are related with
\begin{eqnarray}
b^{t}\equiv B^\mu u_{\mu},\\
b^{i}\equiv (B^i+u^i b^t)/u^t.
\end{eqnarray}
$b^\mu u_{\mu}=0$ is satisfied.
By using this magnetic four vector electromagnetic component
of energy momentum tensor and the dual of Faraday tensor
can be written as
\begin{eqnarray}
T_{\rm EM}^{\mu \nu}=b^2 u^\mu u^\nu + p_{\rm b} g^{\mu \nu}-b^\mu b^\nu,\\
\label{dual_Faraday}
^{*}F^{\mu\nu}=b^{\mu}u^{\nu}-b^{\nu}u^{\mu}
=
{{\mathcal B}^{\mu}u^{\nu}-{\mathcal B}^{\nu}u^{\mu} \over \alpha u^{t}}=
{B^{\mu}u^{\nu}-B^{\nu}u^{\mu} \over u^{t}},
\end{eqnarray}
where the magnetic pressure is $p_{\rm b}=b^\mu b_\mu/2 =b^2/2$.
The Maxwell equations are written as
\begin{eqnarray}
\label{Maxwell_Eq}
{^{*}F^{\mu\nu}}_{;\nu}=0.
\end{eqnarray}
By using Eqs.~(\ref{dual_Faraday})
these equations give
\begin{eqnarray}
\partial_i \left(\sqrt{-g} {B^i}\right)=0,\\
\partial_t \left(\sqrt{-g} {B^i}\right)
+\partial_j \left(\sqrt{-g} (b^i u^j-b^j u^i)\right)=0.
\end{eqnarray}
These are no-monopole constraint equation
and time evolution of spacial magnetic field equations,
i.e, the induction equations, respectively.

In order to close the equations,
ideal gas equation of state $p_{\rm th}=(\gamma-1)U$
is adopted,
where $\gamma$ is the specific heat ratio which is assumed
to be constant ($\gamma=4/3$)
\footnote{The calculation with $\gamma=5/3$
shows some minor differences compared to that
with $\gamma=4/3$ 
as discussed by \citet{McKinney04,Mignone07}.
Similar time variavlities in the inflows and outflows
discussed below are observed
by adopting two different specific heat ratio in the equation of state.}.
We ignore self-gravity of the gas around the black hole
and any radiative processes,
assuming radiatively inefficient accretion flow (RIAF) in the disk
\citep{Narayan94},
although the effects of radiation have been
discussed by \citet{Ryan17}.

We numerically solve these equations by GRMHD code
developed by one of authors \citep{Nagataki09,Nagataki11}.
Magnetohydrodynamic equations are solved
by using shock capturing method (HLL method),
applying 2nd order interpolation to reconstruct of
physical quantities at the cell surfaces
and 2nd order time integration by using TVD (total variation diminishing)
Runge-Kutta method,
see also \citet{Gammie03,Noble06}.
The boundary conditions are zero gradient for $x_1$
and periodic one for $x_3$.

\subsection{Initial Condition}
\begin{figure}
\begin{center}
\rotatebox{0}{\includegraphics[angle=0,scale=0.4]{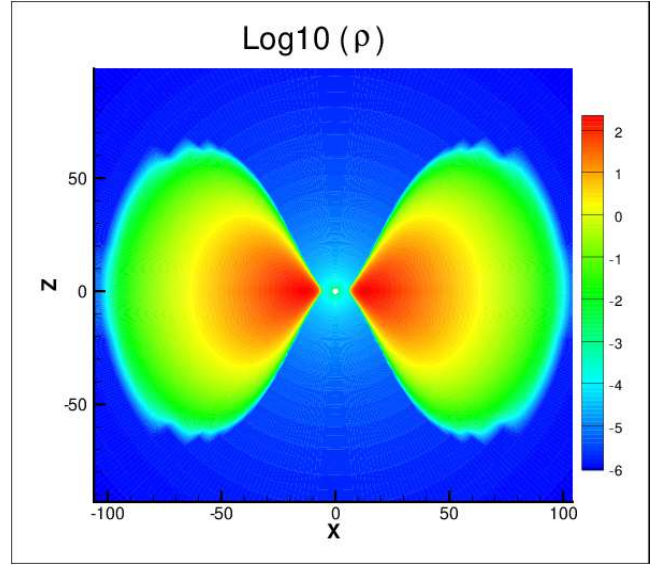}}
\caption{Log scaled rest mass density profile on $x-z$ plane at $t=0$.
\label{initialmass}}
\end{center}
\end{figure}

We adopt the Fishbone-Moncrief solution
as an initial condition for hydrodynamic quantities
as adopted for recent GRMHD simulations
\citep{McKinney04,McKinney06,McKinney12,Shiokawa12}.
This solution describes the hydro-static torus solution
around a rotating black hole.
The gravitational force by the central black hole,
the centrifugal force and the pressure gradient force balance each other.
There are some free parameters to give a solution.
We assume the disk inner edge at the equator is at $r=6.0~R_{\rm g}$ and 
a constant specific angular momentum ($l^{*}\equiv |u^{t}u_{\phi}|=4.45$).
The 4-velocity is firstly given at Boyer Lindquist coordinate then
transformed to Kerr-Schild one and to modified Kerr-Schild one.
The dimensionless spin parameter is assumed to be $a=0.9$.
The radii of event horizon and the innermost stable circular orbits
(ISCO) at the equator are at
$r_{\rm H}(a=0.9)\sim 1.4~R_{\rm g}$ and $r_{\rm ISCO}(a=0.9)\sim 2.32~R_{\rm g}$, respectively.
The initial disk profile is on the plane including polar axis
shown in Fig.~\ref{initialmass}
which shows mass density contour.
The disk is geometrically thick and
is different from standard accretion disk \citep{Shakura73}
in which the geometrically thin disk is assumed.

As we have discussed in Sec.~\ref{sec:intro}
magnetic fields play an important role not only for the dynamics
of the accretion flows but also for the dynamics of the outflows.
We impose initially weak magnetic field inside the disk as a seed.
This weak magnetic field violates the initial static situation
and is expected to be amplified by winding and MRI.

Here we introduce the four-vector potential $\bm A$
of the electromagnetic field.
The Faraday tensor is defined by this vector potential as follows.
\begin{eqnarray}
F_{\mu\nu}=\partial_\mu A_{\nu}-\partial_\nu A_{\mu}.
\end{eqnarray}

In this study we assume initially closed poloidal magnetic field,
i.e., the toroidal component
of the initial vector potential is,
\begin{eqnarray}
\label{mag_initial}
A_\phi \propto \max ( \rho/\rho_{\max}-0.2, 0 ),
\end{eqnarray}
where $\rho_{\max}$ is maximum mass density in the initial torus.
Other spacial components are zero, i.e., $A_r=A_{\theta}=0$.
The minimal plasma $\beta$ which is the ratio of the
thermal pressure to the magnetic pressure
is 100 in the disk.
Since the vector potential has only toroidal components,
the poloidal magnetic field is imposed.
To violate axis-symmetry maximally 5\% amplitude random perturbation
is imposed in the thermal pressure 
i.e., thermal pressure is
$p_{\rm th}= p_0 (0.95+0.1C)$, where
$p_0$ is equilibrium  thermal pressure
derived by Fishbone-Moncrief solution
and  $C$ is random number 
in the range of $0\le C\le 1$.
This perturbation violates axis-symmetry of the system
and triggers generation of non axisymmetric mode.

\section{EPISODIC ERUPTION OF DISKS AND JETS}
\label{results}

At first
we discuss the results based on the highest resolution calculation.
The resolution effect, i.e., comparison with the results
by lower resolution calculations is briefly discussed later.
The main properties, such as the amplification
and the dissipation of the magnetic field inside the disk,
Alfv\'en wave emission from the disk, and these time variavilities,
which will be discussed below
are common for all calculations by different resolutions,
although characteristic timescales are different each other.

\begin{figure*}
\begin{center}
\rotatebox{0}{\includegraphics[angle=270,scale=0.6]{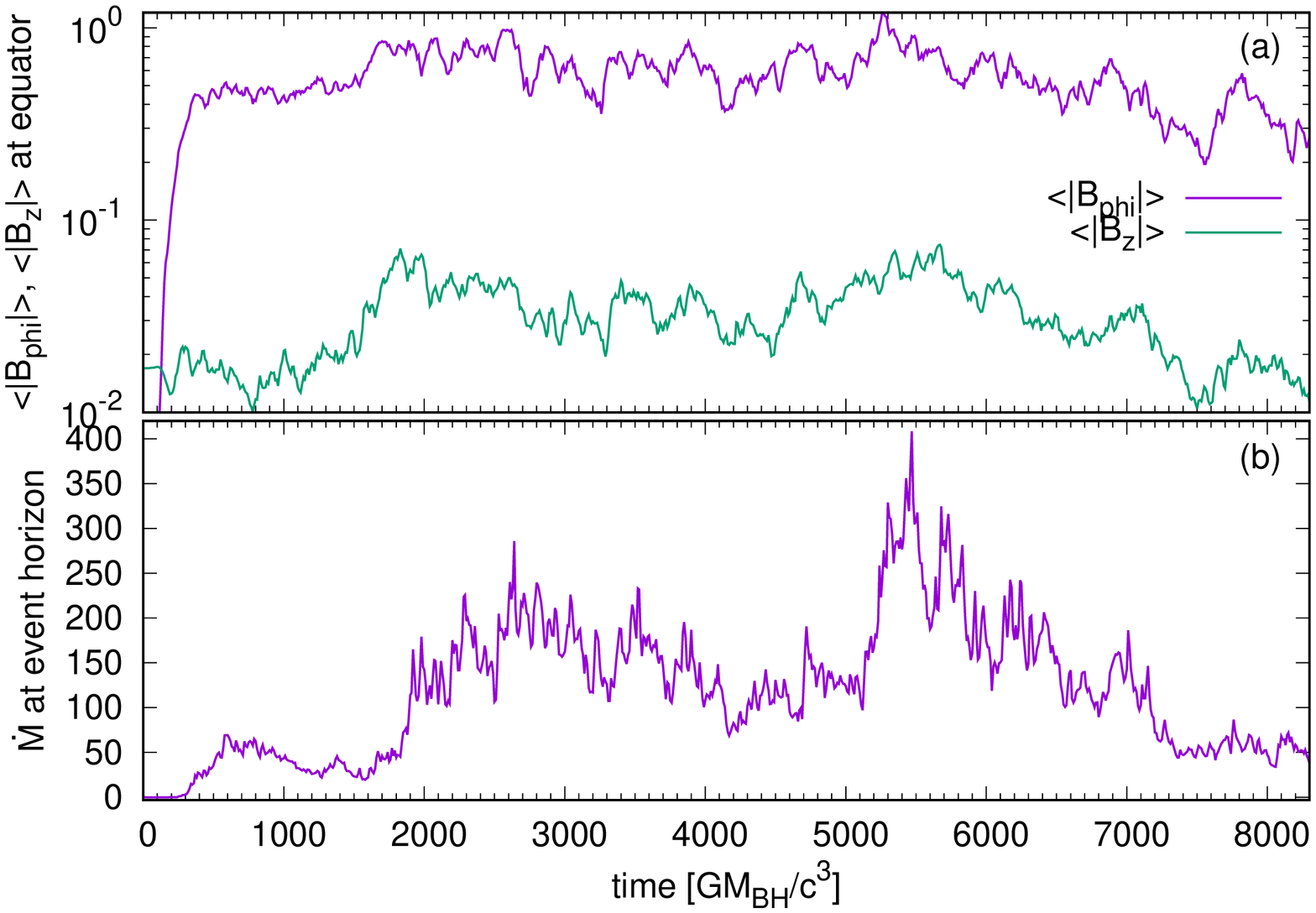}}
\caption{The disk evolution,
(a) time evolution of strength of toroidal (purple) and 
poloidal (green) magnetic fields 
at the equator averaged at $r_{\rm ISCO} \le r \le 10R_{\rm g}$
and $0 \le \phi \le 2\pi$
and
(b) time evolution of mass accretion rate ($\dot M$) at the event
horizon ($r=1.4 R_{\rm g}$).
\label{magden}}
\end{center}
\end{figure*}

\subsection{B-field amplification and mass accretion}
The magnetic field in the disk is amplified by winding effect
and MRI, as follows.
The initially imposed poloidal magnetic field
is stretched in the toroidal direction,
generating toroidal components,
since there is a differential rotation inside
the accretion disk.
Although initially imposed magnetic field is weak,
i.e., the minimum plasma $\beta$ is 100 inside the disk, 
the strength of the magnetic field quickly increases
by the winding effect \citep{Duez06} and MRI.
Soon stratified filaments which is parallel to the equatorial plane 
appear around the equator in the magnetic pressure contour.
Fig.~\ref{magden}(a) shows
the volume averaged
strength of the magnetic field $\langle(B^{i}B_i)^{1/2} \rangle$ 
at the equator i.e., averaged $(B^{i}B_i)^{1/2}$ 
over $r_{\rm ISCO} \le r \le 10R_{\rm g}$, $\theta=\pi/2$,
and $0 \le \phi \le 2\pi$.
Since the magnetic field is stretched by the differential 
rotation in the disk,
toroidal component dominates
over the poloidal component after $t=100~{\rm GM_{BH}}c^{-3}$,
though both components show strong time variability.
Fig.~\ref{magden}(b) shows
the  mass accretion rate at the event
horizon
$\dot M(r_{\rm H}, t)$,
where the mass accretion rate at a radius is defined as
\begin{eqnarray}
\dot M(r_{\rm H}, t)=-\int \sqrt{-g_{\rm KS}} \rho(r_{\rm H},t) u^{r}(r_{\rm H},t) dA_{\rm KS}\nonumber \\
=-\int \sqrt{-g_{\rm mKS}} \rho u^{1} dA_{\rm mKS},
\end{eqnarray}
where $dA$ is area element,
for example, $dA_{\rm mKS}=\Delta x^2 \Delta x^3$, and
the sign is chosen so that the case of mass inflow 
is positive mass accretion.
The mass accretion rate $\dot M(r_{\rm H}, t)$
also shows strong time variability,
and synchronized with 
magnetic field strength near the ISCO.

\begin{figure*}
\begin{tabular}{c|c|c}
Parameter & High $\beta$ disk & Low $\beta$ disk\\ 
\hline
$\beta\equiv$ $P_{\rm gas}/P_{\rm mag}$ &
$\beta>1$ ($P_{\rm gas}$ supported) &
 $\beta\lesssim 1$ ($P_{\rm mag}$ supported) \\
Configuration & 
Optically thick disk &
Optically thin disk \\
 & 
(cooling-dominated) &
(advection-dominated) \\
 & 
 + corona &
consisting of blobs \\
Dissipation of magnetic fields & 
Escape via buoyancy &
Reconnection\\
 & and reconnection&
  \\
Dissipation of energy & 
Continuous &
Sporadic \\
Spectrum & 
Soft + hard tail &
Hard power law \\
Fluctuations & 
 Small & Large \\
\hline
 & & \\
&  \includegraphics[scale=0.33]{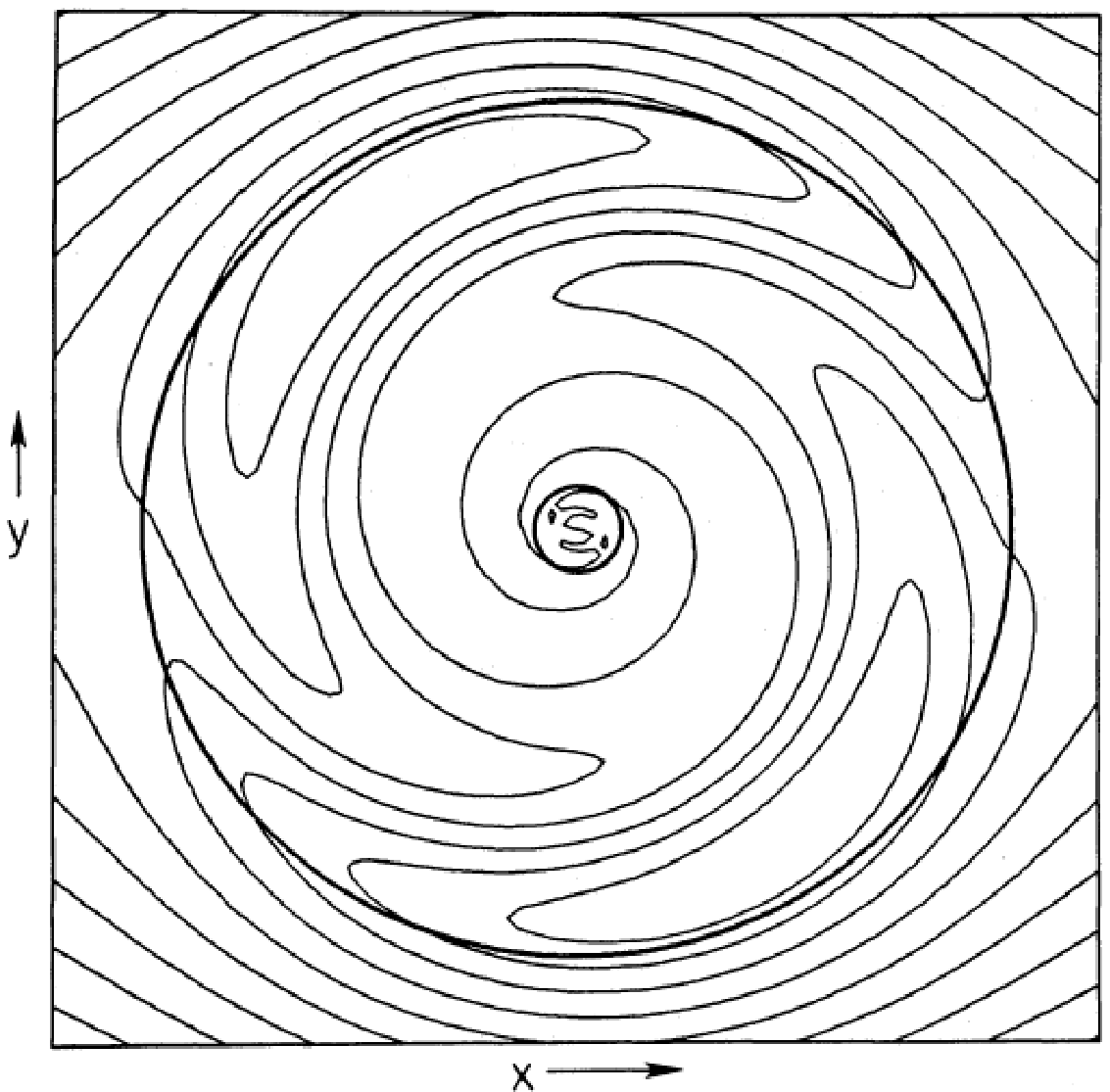}
&  \includegraphics[scale=0.33]{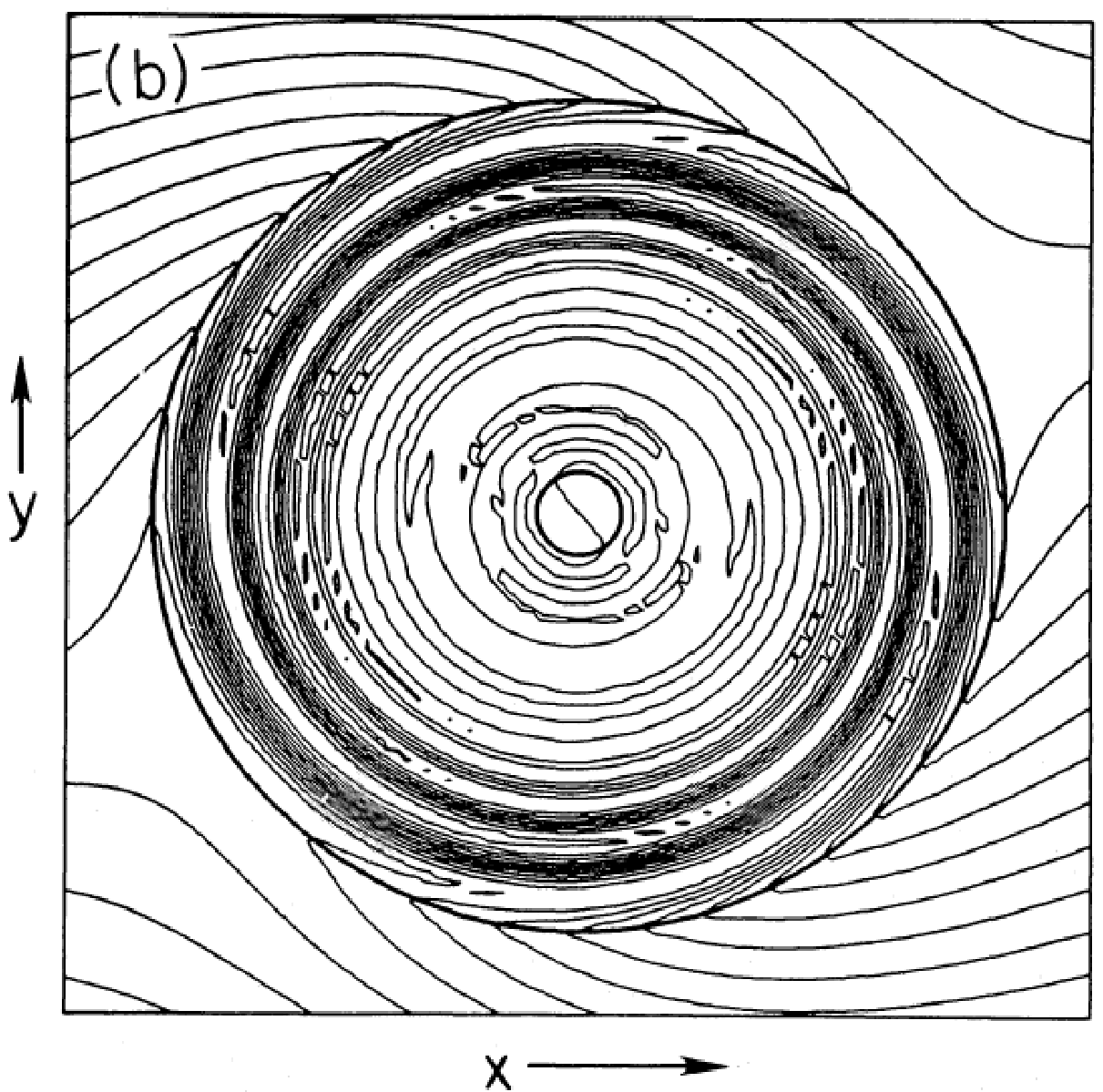}
\\
&
 \includegraphics[scale=0.25]{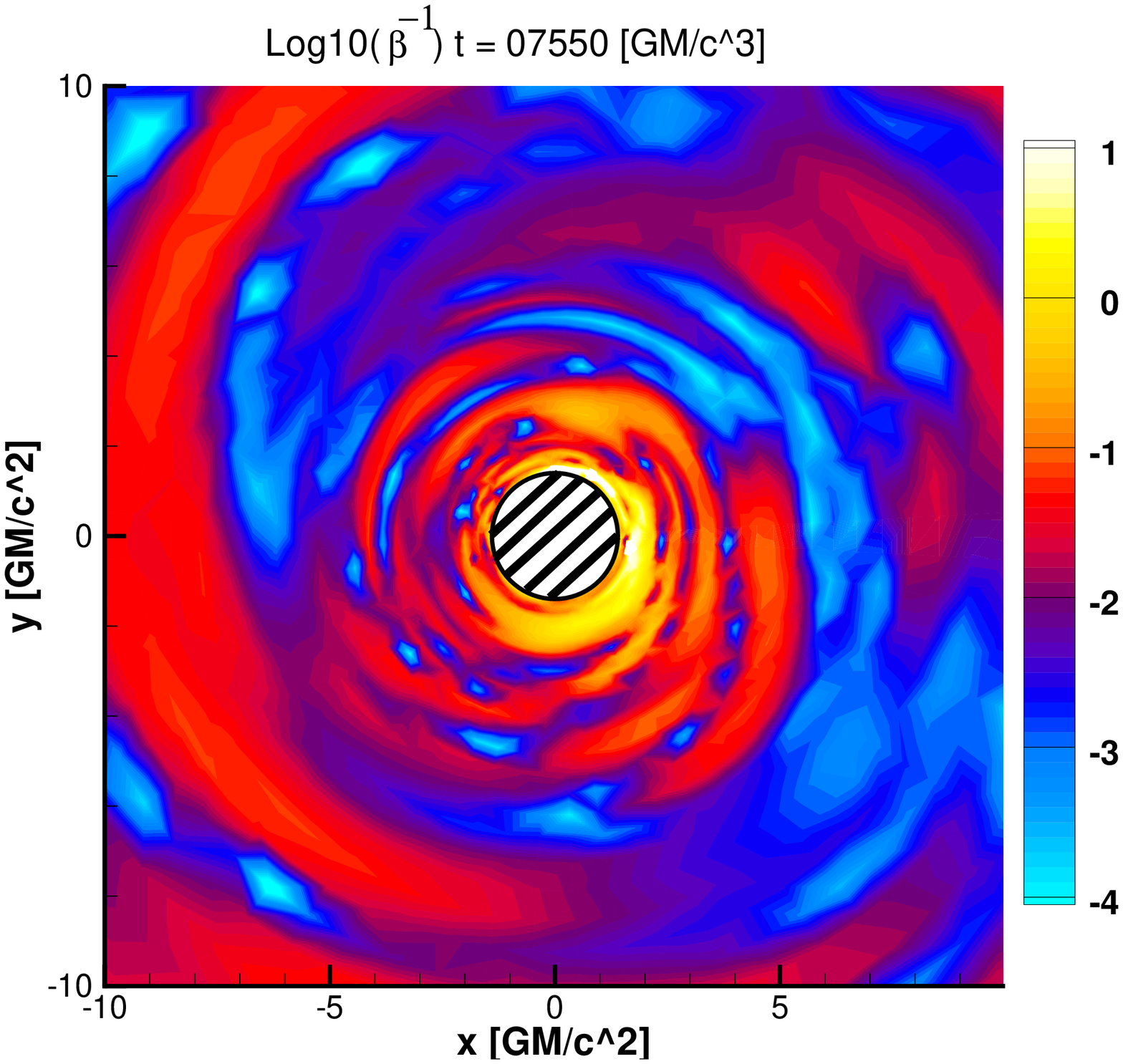}
&  \includegraphics[scale=0.25]{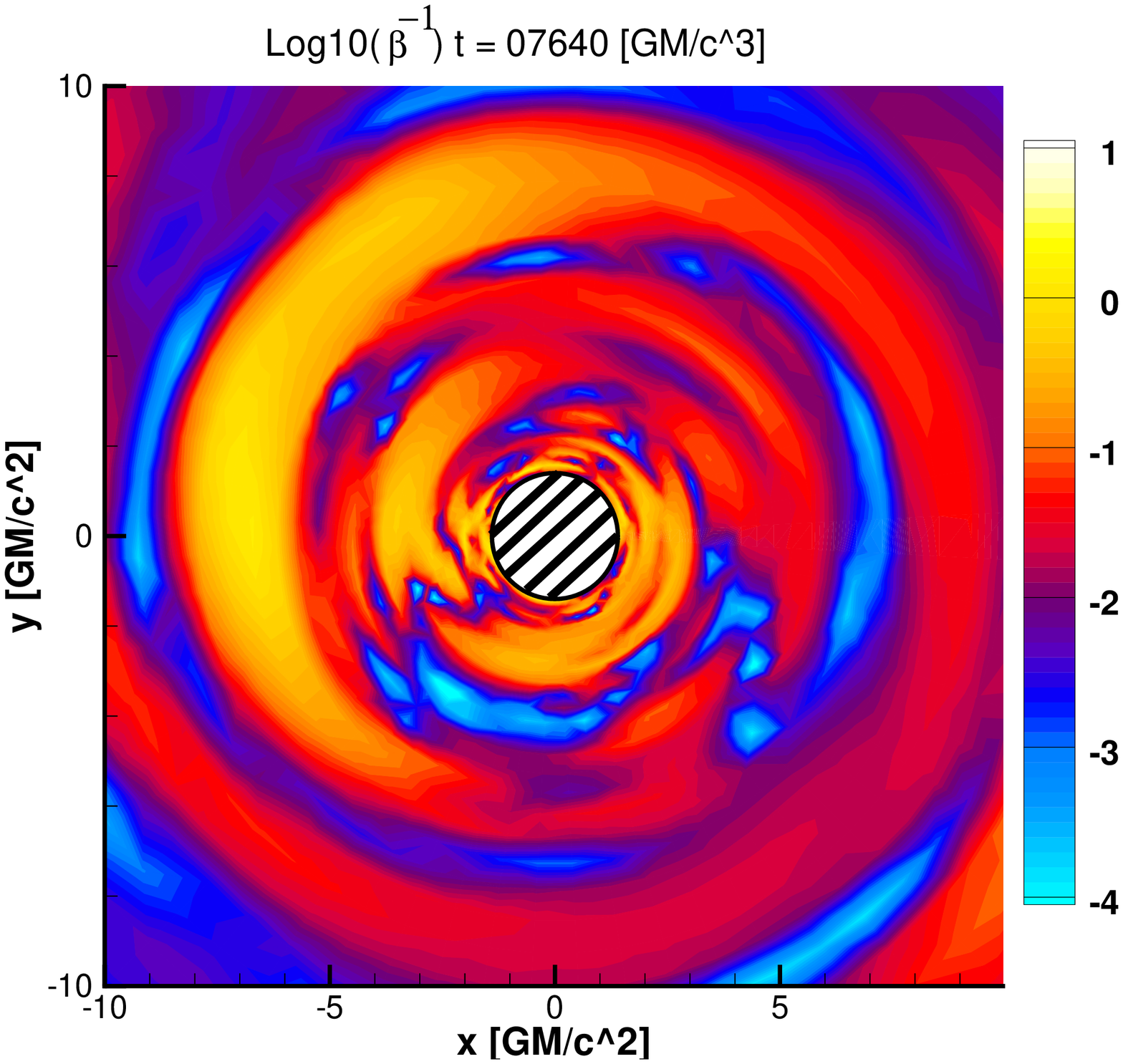}
\end{tabular}
\caption{Table: Properties of high and low $\beta$ states of the disks taken
 from \citet{Mineshige95} (\copyright AAS. Reproduced with permission).
Middle panels : Toroidal magnetic field lines at high $\beta$ state
(left) and low $\beta$ state taken from \citet{Tajima87}
(\copyright AAS. Reproduced with permission).
Bottom panels : Inverse of plasma beta ($\beta^{-1}$) contours at the
 equator shown by logarithmic scales 
at $t=7550 {\rm GM_{\rm BH}}c^{-3}$
(low $\beta$  state, left)
and at $t=7640 {\rm GM_{\rm BH}}c^{-3}$ (low $\beta$ state, right).
The shadowed regions in the circle indicate inside the event horizon.
\label{plasmabeta_TEST600}}
\end{figure*}

Bottom two panels in Fig.~\ref{plasmabeta_TEST600}
show the contours of $1/\beta$ at the equator
at $t=7550~{\rm GM_{\rm BH}}c^{-3}$ and $t=7640~{\rm GM_{\rm BH}}c^{-3}$.
Between these two figures
the state of the disk near the disk inner edge ($r \sim 6 R_{\rm g}$)
transitioned from high $\beta$ state ($\beta^{-1} \sim 10^{-2}$)
to low $\beta$ state ($\beta^{-1} \sim 1$).
Since we sometimes
observe that the plasma $\beta$ in the disk is more than 100,
we use ``low'' $\beta$ state for the disk
with a plasma $\beta$ of order of unity
at which the disk is still gas pressure supported.
It should be noted that
``low $\beta$ state'' is defined as a magnetically supported disk
with $\beta \lesssim 1$, for example, \citet{Mineshige95}.
The characteristics of two states presented in \citet{Mineshige95}
are listed in the top of Fig.~\ref{plasmabeta_TEST600}.
Middle two panels in Fig.~\ref{plasmabeta_TEST600}
show the toroidal magnetic field lines
of two disks from MHD simulations
by \citet{Tajima87}.
At the low $\beta$ state (right panels in
Fig.~\ref{plasmabeta_TEST600}) the toroidal magnetic field
is stretched to the limit
and magnetic field energy is stored.
Since some field lines are almost anti-parallel
and very close to each other,
the reconnection will happen eventually.
After magnetic field energy is dissipated
via reconnection,
the system becomes high $\beta$ state
(left panels in Fig.~\ref{plasmabeta_TEST600}).

\begin{figure*}
\begin{center}
\includegraphics[angle=0,scale=0.4]{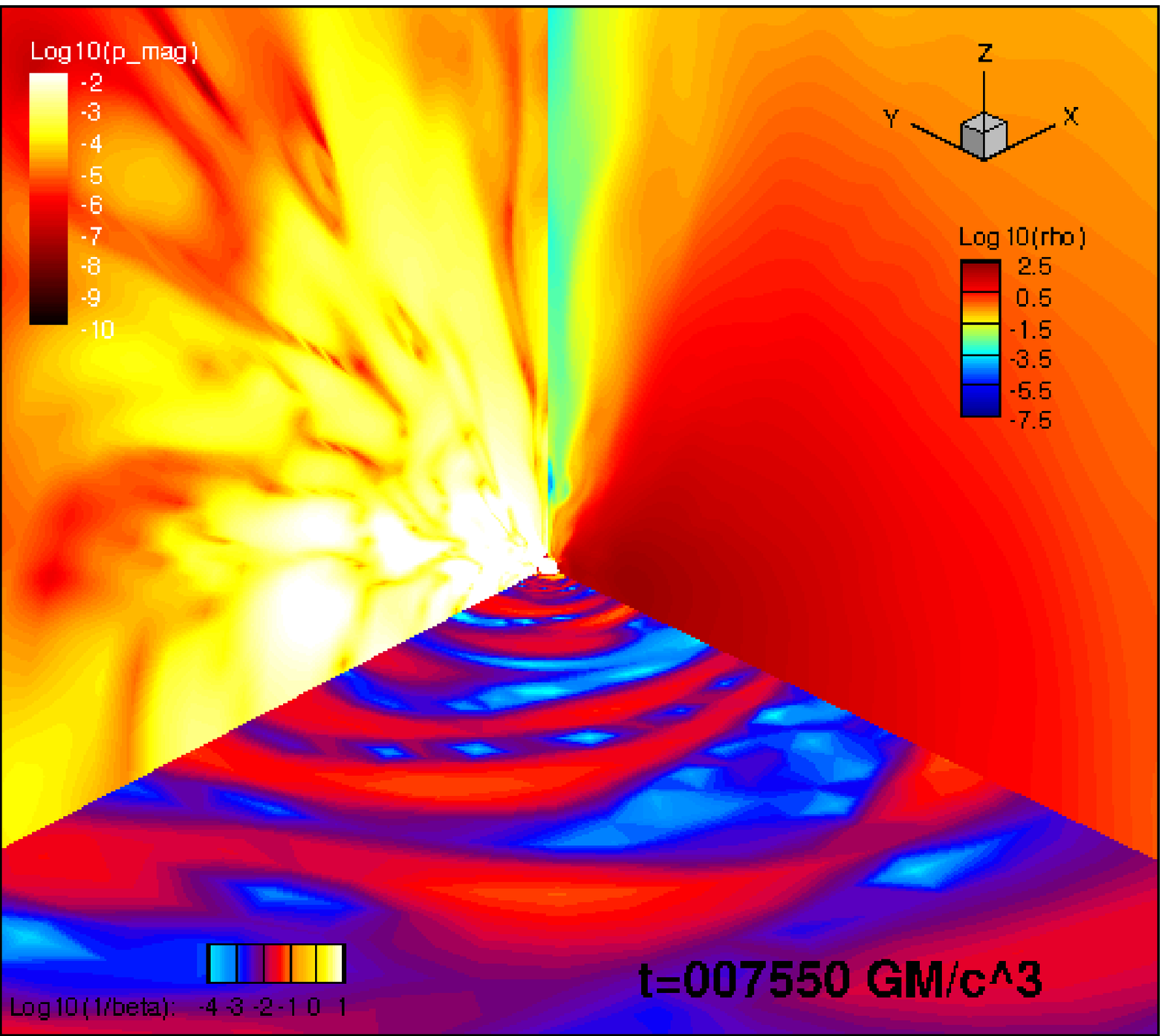}
\includegraphics[angle=0,scale=0.4]{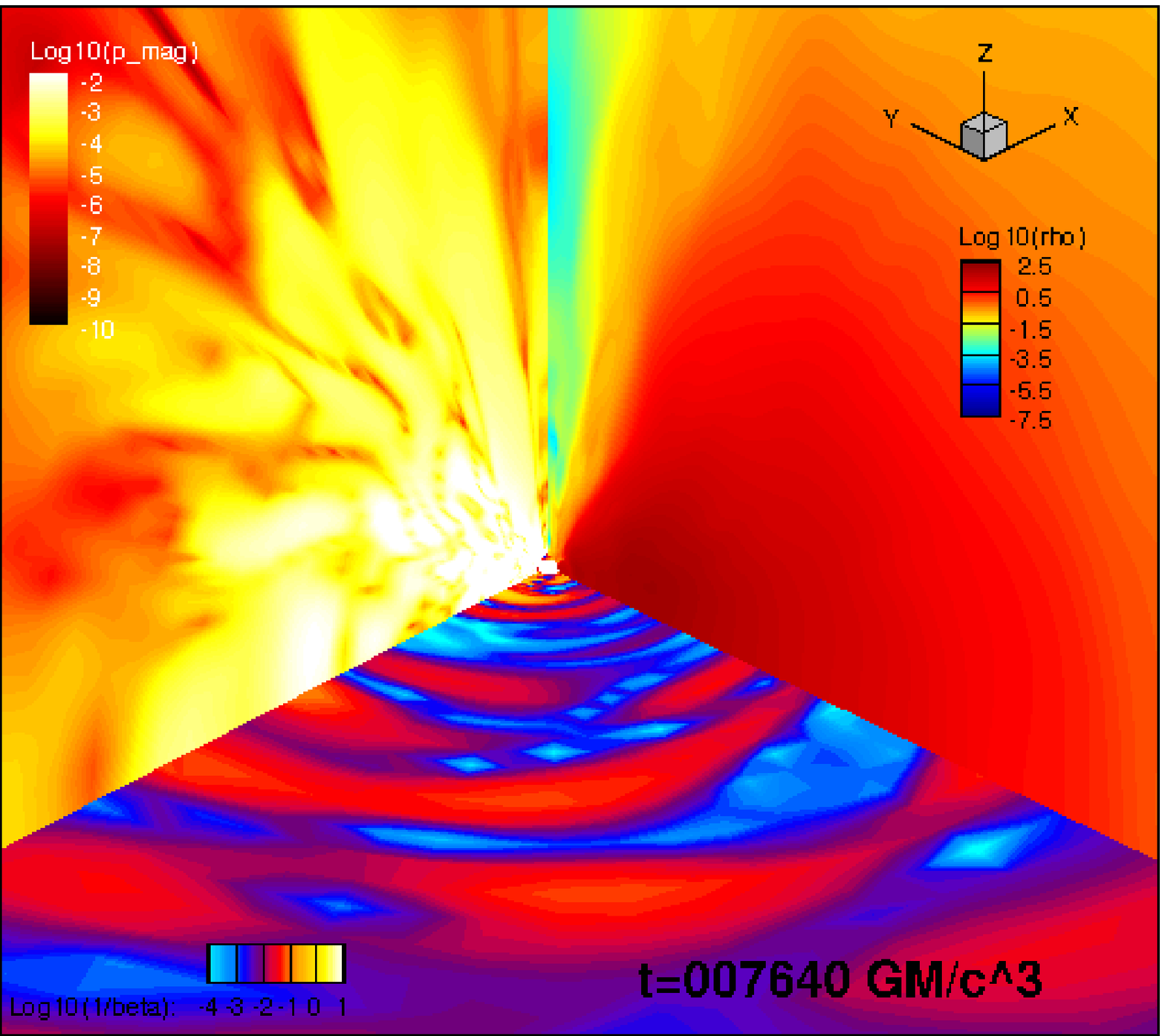}
\caption{Contours of inverse of the plasma $\beta$ (at the equator), 
mass density ($y-z$ plane), and magnetic pressure ($x-z$ plane)
at two different times as shown in Fig.~\ref{plasmabeta_TEST600}.
The domain shows $-80 {\rm GM_{BH}} c^{-2}< x <0$, $-80 {\rm GM_{BH}} c^{-2}< y <0$, and
$0< z <70 {\rm GM_{BH}} c^{-2}$.
\label{panels}}
\end{center}
\end{figure*}

Bar structures near the disk inner edge can be seen
(bottom panels in Fig.~\ref{plasmabeta_TEST600}).
The non-axis-symmetric mode is excited,
as shown in global hydrodynamic and magnetohydrodynamic
simulations of accretion disks,
for example \citet{Tajima87,Machida03,Kiuchi11,McKinney12}.
Figure~\ref{panels} shows the contours of $1/\beta$
(at the equator), 
mass density ($y-z$ plane), and magnetic pressure ($x-z$ plane)
at two different times as shown in Fig.~\ref{plasmabeta_TEST600}.
Along the polar axis low density and highly magnetized region
appears.
It corresponds to the Poynting flux dominated jet.
Thus baryonless and highly magnetized jet is
formed along the polar axis.
In this region the Alfv\'en speed is almost speed of light $\sim c$. 
A disk wind blows between the jet and the accretion disk.
Filamentaly structures which are excited by MRI can be seen
in the magnetic pressure contours.
The thickness of filaments is $\sim 0.1 R_{\rm g}$,
as shown in Fig.~\ref{panels}.

Both the averaged strength of the magnetic field near the ISCO
at the equator and the mass accretion rate at the event horizon
show synchronous time variability.
This is because that the mass accretion rate at the event horizon
is strongly affected by the activities of magnetic field
amplification near the disk inner edge.
The magnetic field amplification via MRI
enhances the specific angular momentum transfer inside the accretion disk,
resulting in the increase of the accretion rate
at the event horizon.
This means that the amplification of magnetic fields
acts as a viscosity which is introduced as $\alpha$-viscosity
in \citet{Shakura73}.
Typical increase timescales are $20-60 ~GM_{\rm BH} c^{-3}$.
While the magnetic field is amplified,
the mass accretion rate at the event horizon rapidly increases.
As we will show later,
the outflow properties are also show intense time variability
which is strongly related with particle acceleration via
wakefield acceleration.

The mass accretion rate repeatedly shows sharp rises followed by
gradual falling down.
The rising timescale for the quick increase of mass accretion rate
corresponds to the value of
the growth timescale of MRI
\begin{eqnarray}
\tau_1 \sim f_{\rm MRI}{\Omega_{\rm MRI}}^{-1},
\end{eqnarray}
where $f_{\rm MRI}$ is order of unity,
$\Omega_{\rm MRI}$ is growth rate of MRI, and
$\Omega_{\rm MRI}=3\Omega_{\rm K}/4$ for the fastest growing mode
and $\Omega_{\rm K}(r)=r^{-3/2}$ is Newtonian Keplerian angular
velocity.
The timescale for the fastest growing MRI mode is
\begin{eqnarray}
\tau_1 \sim 4.7 f_{\rm MRI}
\left({r\over r_{\rm ISCO}}\right)^{3/2}[{\rm GM}_{\rm BH} c^{-3}].
\end{eqnarray}
This timescale at around $r= 6-8 ~R_{\rm g}$
is almost as same as the timescale
of increase of the strength of magnetic field in the disk.
By the analysis of MRI for Newtonian MHD,
the angular frequency of the mode is
$k_z {{c_{\rm A}}_z}=\sqrt{15/16}\Omega_{\rm K}\sim \Omega_{\rm K}$ 
for the fastest growing mode for $z$ (parallel to polar axis) direction
in Keplerian accretion disk,
where $k$ is wave number and ${{c_{\rm A}}_z}$ is Alfv\'en speed
of $z$ component.
The volume averaged Alfv\'en speed $\langle {c_{\rm
A}}_{z}\rangle
\sim \langle \sqrt{b^{\theta} b_{\theta}/(\rho h +b^2)} \rangle$
near the ISCO ($r_{\rm ISCO}<r<10 ~R_{\rm g}$) at the equator
is typically $3\times 10^{-3}~c$.
Thus the wavelength of this mode is 
$\lambda=2\pi/k_z\sim
 2\pi \langle {c_{\rm A}}_{z}\rangle /\Omega_{\rm K}\sim 0.022 (r/r_{\rm
 ISCO})^{1.5} R_{\rm g}$
with a grid size near the ISCO and at the equator
$r\Delta \theta=0.0057(r/r_{\rm ISCO})R_{\rm g}$ for higher resolution case.
Our simulation shows that
the typical rising timescale in poloidal magnetic field amplification
$\sim 30 {\rm GM_{\rm BH}} c^{-3}$.
Corresponding wavelength of MRI is estimated $\sim 0.14R_{\rm g}$
at $r=8 R_{\rm g}$.
This structure is well resolved by more than $8$ grids 
and is
consistent with thickness of the filamentary structure near the equator
shown in magnetic pressure panel in Fig.~\ref{panels}.

Since the episodic period of this quick increase
of strength of poloidal magnetic field,
i.e., large peak to peak,
is $\tau_2 \sim 100-400{\rm GM_{\rm BH}} c^{-3}$
which is about 2-6 times longer than
the Keplerian orbital period at $r=6 {R_{\rm g}}$
near the ISCO
($\sim 22 (r/r_{\rm ISCO}){\rm GM_{\rm BH}} c^{-3}$).
This timescale for the recurrence 
is roughly consistent with
the analysis by
local shearing box \citep{Stone96,Suzuki09,Shi10,O'Neill11}
in which about 10 times orbital period at the radius of magnetic field
amplfication is observed as repeat timescale.

Along the polar axis funnel nozzles appear (Fig.~\ref{panels}).
The outward going electromagnetic luminosity, opening angle of this jet
are also time variable like as the mass accretion rate.
The radial velocity just above the black hole
and becomes positive at typically
$10\lesssim r \lesssim 20 R_{\rm g}$,
i.e., stagnation surfaces.

\begin{figure*}
\begin{center}
\rotatebox{0}{\includegraphics[angle=270,scale=0.8]{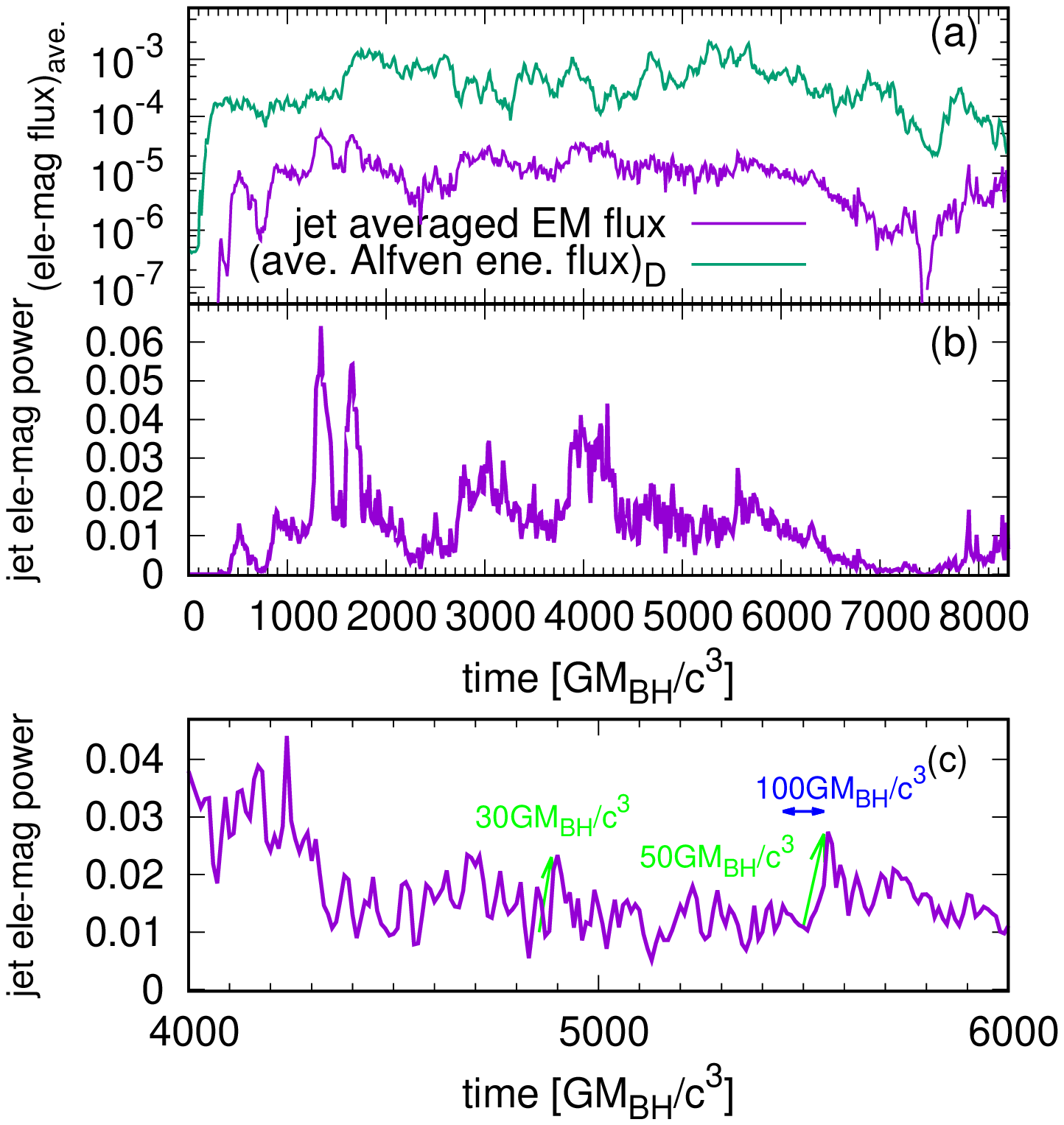}}
\caption{
The jet activities.
(a): time evolution of area averaged radial electromagnetic flux
at $r=15 {R_{\rm g}}$ and $0<\theta<20^{\circ}$ (purple).
Time evolution of averaged electromagnetic flux inside
the disk calculated
by $(\langle E_{\rm EM}\rangle/\langle dV \rangle) 
\langle {c_{\rm A}}_{z}\rangle /2$
at the equator and at $r_{\rm ISCO}<r<10 {R_{\rm g}}$ (green).
(b) Time evolution of
electromagnetic (Poynting) power in the jet at $r=15 R_g$
calculated by 
the area integration of electromagnetic flux at $0<\theta<20^{\circ}$.
(c) same as (b) but
the period from $t=4000 GM_{\rm BH} c^{-3}$ to $t=6000 GM_{\rm BH} c^{-3}$
is shown.
Rising and repeat timescales of flares
are presented in the figure.
\label{ene_horizon}}
\end{center}
\end{figure*}

Figure~\ref{ene_horizon}(b) shows radial electromagnetic luminosity
calculated by the area integration only around polar region
($0\le\theta \le 20^{\circ}$)
at the radius $r=15 {\rm R_g}$.
Electromagnetic luminosity shows similar short time variability
with the magnetic field amplification near the disk inner edge.
We can see typical rising timescale of the flares
is as same as that for rising timescale of magnetic field
in the disk, i.e., $\bar \tau_1 \sim 30{\rm GM_{BH}} c^{-3}$
and
the typical cycle of flares is also as same as repeat cycle of
magnetic field amplification
$\bar \tau_2 \sim 100{\rm GM_{BH}} c^{-3}$.
We have observed some active phases in electromagnetic luminosity in
the jet.
In these active phase,
the averaged radial electromagnetic flux
increases and becomes about a few tens percent of the averaged
disk Alfv\'en flux at the equator at around
$t=1300, 3000, 4000,$ and $8200 {\rm GM_{BH}}c^{-3}$,
as shown in the
Fig.~\ref{ene_horizon} (a).
The disk Alfv\'en flux at the equator is evaluated
as an average of $z$ component of Alfv\'en energy flux at
the equator  $\langle E_{\rm EM}/dV\rangle$
times half of  Alfv\'en speed $\langle {c_{\rm A}}_z\rangle/2$
inside the disk ($r_{\rm ISCO}<r<10 R_{\rm g}$).
A large fraction of emitted Alfv\'en waves goes into the jet,
when the level of the electromagnetic luminosity in
the jet becomes a few tens percent of Alfv\'en flux in the disk.
This is almost consistent with the assumption A
as we discuss in the next section.

\begin{figure*}
\begin{center}
\rotatebox{270}
{\includegraphics[angle=0,scale=0.7]
{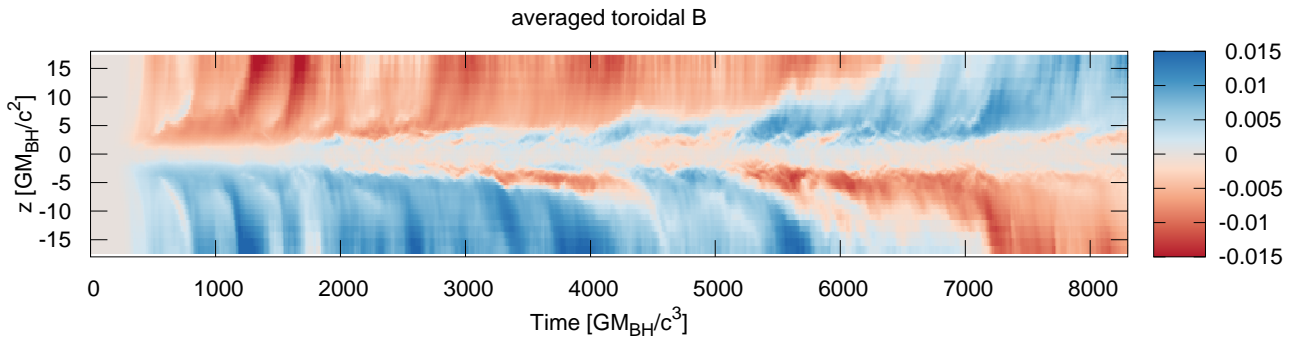}}
\caption{
The time evolution and vertical structure of the
averaged toroidal magnetic field
at $0\leq \phi \leq 2\pi$
and $3R_{\rm g}\leq R\leq 3.2R_{\rm g}$, where $R$ is distance
from the polar axis.
}
\label{butterfly}
\end{center}
\end{figure*}

Figure~\ref{butterfly} shows time evolution and the vertical structure of the
averaged toroidal magnetic field
($\langle(B^{\phi}B_{\phi})^{1/2} \rangle$)
as shown in \citet{Shi10,Machida13}, i.e, butterfly diagram.
The average is taken  at $0\leq \phi \leq 2\pi$
and $3R_{\rm g}\leq R\leq 3.2R_{\rm g}$, where $R$ is distance
from the polar axis.
Although initially no magnetic field is imposed at $R=3R_{\rm g}$,
soon the magnetic field is transported to this site
with accreting gas due to MRI growth and angular momentum exchange.
After that the magnetic fields quasi-periodically goes up to
the north and goes down to the south from above and below the
equator due to the magnetic buoyancy,
i.e., Parker instability \citep{Parker66},
as shown by \citet{Suzuki09,Shi10,Machida13}.
Each episode corresponds to the one cycle 
of the disk state transitions.
The magnetic fields sometimes changes its sign,
although it happens much less than that observed
in \citet{Shi10}
who did high resolution simulations in local hearing box.
Around the equator
magnetic fields rise up with a speed
$\sim 10^{-3}~c$ which corresponds to the 
averaged Alfv\'en speed of $z$ component (${{c_{\rm A}}_z}$) there.
Strong magnetic fields sometimes goes up or goes down 
from the equator to outside of the disk.
The appearance of the flare in the Poynting luminosity
(Fig.~\ref{ene_horizon} (b) and (c))
in the jet corresponds to this strong magnetic field escape
from the disk.

{The magnetic field lines in the jets are
connected not with the disks but with the 
middle and high latitude of the central black hole.
The outward going electric magnetic luminosity from the 
middle and high latitude of the central black hole
is not so high as compared with that in the jet. 
The  Alfv\'en waves emitted from the disks
do not directly goes into the jets
as assumed in \citet{Ebisuzaki14}.
As shown above the time variavilities
of the Poynting flux in the jet is as same as
that of the magnetic fields strength in the disk
and shows strong correlation.
The amplification of Poynting flux 
above the black hole and the disks occurs
by the Alfv\'en fluxes from the disk.
Another possibility is
that the blobs falling onto the black hall 
interact with magnetic fields which are connected with
those in the jets.

\subsection{resolution effect}
In this subsection we discuss the resolution effect.
We have performed calculations by
three different grid types in polar and azimuthal angle
as described in sec. \ref{sec:basic_eq2}.
The highest resolution case for which
the results are shown above
is by non-uniform grids in polar angle
for which the grids concentrates around the equator,
resulting in about 10 or 5 times better in the polar angle
around the equator than that for other two cases
in which constant polar angle grids are adopted with around a half or
same number of polar grids.
In all cases
we have observed properties discussed above,
such as,
time variable magnetic field amplification in the disk,
disk state transition between low and high plasma $\beta$ states,
time variable mass accretion onto black holes,
and Poynting flux dominated jet with some flares. 
The timescale of the fastest growing mode 
($30 {\rm GM_{BH}}c^{-3}$) is observed
in the amplification of magnetic fields in the disk
for highest resolution case,
whereas longer timescales (typically $50 {\rm GM_{BH}}c^{-3}$
for 2nd highest resolution case 
and $80 {\rm GM_{BH}}c^{-3}$ or longer time scale for
the lowest resolution case)
are observed.
The thinnest and multiple filaments 
in magnetic pressure contour around the equator
are observed for the highest resolution case.
These results mean the longer wavelength mode of MRI
is observed in the lower resolution cases.
The recurrence timescale for higher resolution case
is also faster than that by low resolution cases.

\section{Particle Acceleration}
\label{sec:discussion}
As shown in Fig.~\ref{ene_horizon} (b)
flares in electromagnetic power in the jet are
observed,
where the Alfv\'en speed is almost speed of light
because of the low mass density.
Large amplitude Alfv\'en waves become
electromagnetic waves by
mode conversion
of strongly relativistic waves \citep{Daniel97,Daniel98,Ebisuzaki14}.
The interaction of the electromagnetic waves and the plasma
can result in the acceleration of the charged particles
by the ponderomotive force
i.e., wakefield acceleration \citep{Tajima79}.
The key for the efficient
wakefield acceleration is
the Lorentz invariant dimensionless strength parameter of the wave
\citep{Esarey09},
\begin{eqnarray}
a_0={eE \over m_e \omega c}.
\end{eqnarray}
The velocity of the oscillation motion of the charged particles
via electric field becomes speed of light,
when $a_0 \sim 1$.
If the strength parameter $a_0$ highly exceeds unity,
the ponderomotive force works
to accelerate the charged particles to relativistic regime
to wave propagating direction.

\subsection{Comparison with \citet{Ebisuzaki14}}
In order to evaluate $a_0$,
\citet{Ebisuzaki14} used three assumptions
A, B and C.
Based on the results of numerical simulation,
we intend to confirm three assumptions.
First, assumption A tells us that
the Alfv\'en flux in the jet is equal to that in the disk.
As shown in Fig.~\ref{ene_horizon}(a),
electromagnetic flux in the jet
becomes a few tens percent of Alfv\'en flux in the accretion disk
at the some active phases of the electric magnetic luminosity
in the jet.
Thus most Alfv\'en waves emitted from the disk via Alfv\'en
burst goes to the jet as assumed in \citet{Ebisuzaki14}
at this epoch,
in other words, assumption A.
in which all Alfv\'en waves are assumed to go into the jet.

Second,
\citet{Ebisuzaki14} assumed that magnetic field amplification
occurs at $R=10 {\rm R_s}=20 {\rm R_g}$ for
the standard disk model \citep{Shakura73}.
In \citet{Ebisuzaki14}
the timescale of the magnetic field amplification
($\tau_1$) 
and
frequency of the Alfv\'en wave are
determined by the  MRI growth rate (Eq.~(\ref{growthrate})).
Although they evaluated it around $10~R_{\rm s}$,
the magnetic field amplification occurs
at any radius in the disk.
Since magnetic field amplification which affects to the
mass accretion rate and the time variavilities in the jet
mainly
occurs inside compared with the assumption by \citet{Ebisuzaki14},
the timescales are shorter than those of them due to
faster rotation period.
If we apply $R=6.4{\rm R_g}=3.2{\rm R_s}$ instead of $R=20{\rm R_g}=10{\rm R_s}$
for \citet{Ebisuzaki14} model,
the timescales are
close to our numerical results 
as shown in table \ref{comparisontable}.
The reason why the magnetic field amplification
at smaller radius 
may be due to the high spinning
of the black hole, i.e., $a=0.9$ for which 
both the event horizon and the radius of ISCO
is smaller than those for non-spinning black hole case
($r_{\rm ISCO}(a=0)= 6 ~R_{\rm g}$).
This timescale is well consistent to the rising timescales
of blazar flares observed for 3C454.3
which will be discussed in next subsection.
In other words, assumption B is OK
in qualitatively,
but Eq.~(\ref{growthrate}) must to be $\omega_A\sim 1.0\times10^{-4}(m/10^8)^{-1}{\rm Hz}$~
 (see also table ~\ref{comparisontable2}).

Finally, \citet{Ebisuzaki14}
estimated the repeat timescale as the crossing time of the
Alfv\'en wave in the disk, i.e, $Z_{\rm D}/ {c_{\rm A}}_z$
(assumption C)
for the standard disk model \citep{Shakura73}.
When we apply the radius $R=6.4{\rm R_g}=3.2{\rm R_s}$ instead of
$R=20{\rm R_g}=10{\rm R_s}$ in Eq.~(\ref{recurrencerate}),
we obtain $Z_{\rm D}/ {c_{\rm A}}_z=354 GM_{\rm BH} c^{-3}$,
ignoring factor $\eta$ which is order of unity.
This value is close to our typical $\bar \tau_2=100 GM_{\rm
BH} c^{-3}$.
The case of $R=20{\rm R_g}=10{\rm R_s}$ 
is also listed in the table \ref{comparisontable}.

We can reevaluate the strength parameter $a_0$ as
\begin{eqnarray}
%a_0={eE\over m_e \omega_{\rm A}c}=8.8\times 10^{10} 
%a_0={eE\over m_e \omega_{\rm A}c}=1.4\times 10^{11} 
a_0={eE\over m_e \omega c}=1.4\times 10^{11} 
\left({M_{\rm BH}\over 10^8 M_{\odot}}\right)^{1/2}
\left({\dot M_{\rm av}c^2 \over 0.1 L_{\rm Ed}}\right)^{1/2}.
\end{eqnarray}
Here electric field is estimated as
$E=({\langle{c_{\rm A}}_{\rm D}\rangle/ c})^{1/2}\langle B_{\rm D}\rangle $.
The angular frequency of the
pulsed electromagnetic wave originating from the Alfv\'en shock
(see \citet{Ebisuzaki14})
${\omega}_{\rm D}=2\pi {c_{\rm A}}_{\rm D}/{\lambda_{\rm
A}}_{\rm D}$, where
${\lambda_{\rm A}}_{\rm D}$ is assumed to be 0.14 $R_{\rm g}$.
We use the values at $t=7900GM_{\rm BH} c^{-3}$
and 
the time averaged mass accretion rate 
$7750 GM_{\rm BH} c^{-3}\leq t \leq 8300 GM_{\rm BH} c^{-3}$)
at the event horizon $\dot M_{\rm av}=55.8$ is used as a normalization
to 10\% Eddington luminosity
for $M_{\rm BH}=10^8$ solar masses.
The estimated strength parameter highly exceeds unity
as discussed in \citet{Ebisuzaki14}.
This suggests efficient particle acceleration via wakefield acceleration
can occur in the jet.
Since both electrons and protons are accelerated 
at the large amplitude electromagnetic flares
via ponderomotive forces and these particle
move with the waves,
high energy non-thermal electrons are concentrated
at these waves.

If we apply the radius for the
magnetic field amplification
at $R=6.4{\rm R_g}=3.2{\rm R_s}$ instead of
$R=20{\rm R_g}=10{\rm R_s}$ for Eq.~(\ref{recurrencerate}),
the estimated values such as 
the angular frequency of Alfv\'en wave in the jet
(${\omega_{J}}$),
recurrence rate of the burst ($1/\nu _{\rm A}$),
acceleration time $D_3/c$,
maximum energy of accelerated particle $W_{\rm max}$,
total accretion power $L_{\rm tot}$,
Alfv\'en luminosity $L_{\rm A}$ in the jet,
gamma-ray luminosity $L_{\rm \gamma}$, and
UHECR luminosity $L_{\rm UHECR}$ in \citet{Ebisuzaki14}
are revised.
Table \ref{comparisontable2} is revised version
of Table 1 in \citet{Ebisuzaki14}.
Figure~\ref{MBH_Ltot} is the
revised version of Fig.~4 in \citet{Ebisuzaki14}
which shows the relation of
the maximum energy of accelerated particle $W_{\rm max}$
as a function of mass of central black hole
and accretion power $L_{\rm tot}=\dot M c^2$.
Since we do not consider any radiative processes,
i.e., RIAF ($L_{\rm tot} \le 10\%$ Eddington luminosity),
the gray shadowed region shows the objects
for the UHECR ($W_{\rm max}\ge 10^{20}$ eV) accelerators.

\begin{table*}
\caption{Comparison the timescales unit in $GM_{\rm BH} c^{-3}$
of rising flares and repeat cycle of flares 
between our numerical results, 
blazar observations,
and \citet{Ebisuzaki14} (ET14).
Black hole masses
$M_{\rm BH(3C454.3)}=5\times 10^8M_{\odot}$ \citep{Bonnoli11},
and $M_{\rm BH(AO0235+164)}=5.85 \times 10^8 M_{\odot}$ \citep{Liu06}, 
are used.
For \citet{Ebisuzaki14} model,
two different radii ($R=6.4{\rm R_g}$ and $R=20{\rm R_g}$ (their original assumption))
at which magnetic field amplification occurs are considered.
\label{comparisontable}}
\begin{tabular}{cc|cc|cc}
 & our results 
& 3C454.3 & AO0235+164 & ET14 & ET14  \\
 &             & &  & $R=8{\rm R_g}$ & $R=20{\rm R_g}$   \\
\hline \hline
rising timescale of flares ($\bar \tau_1$) & 30 & $57^{a}$ & $325^{b}$ & 128 &
		     $5.1\times10^2$ \\
repeat cycle of flares ($\bar \tau_2$) & 100 & $132^{a}$ & $433^{b}$ & 354 &
		     $2.0\times 10^3$ \\
\end{tabular}\\
a.~\citet{Abdo11}\\
b.~\citet{Abdo10a}
\end{table*}

\begin{table}
\begin{center}
\caption{Revised version of the Table 1 in \citet{Ebisuzaki14},
where angular frequency in the jet is ${\omega_A}$,
recurrence rate of the burst is $\nu_{\rm A}$,
acceleration length is $D_3$,
maximum energy of accelerated particle is $W_{\rm max}$,
$z$ is the charge of the particle,
$\Gamma$ is bulk Lorentz factor of the jet,
total accretion power is $L_{\rm tot}$,
Alfv\'en luminosity in the jet is $L_{\rm A}$,
gamma-ray luminosity is $L_{\rm \gamma}$, and
UHECR luminosity is $L_{\rm UHECR}$.
$R=6.4 R_{\rm g}$ instead of $R=20 R_{\rm g}$ is assumed
as a radius where the magnetic field amplification occurs.
\label{comparisontable2}}
\begin{tabular}{lll}
 & Values & Units\\
\hline \hline
$2\pi/{\omega_A}_{\rm J}$& $1.2\times 10^2 (\dot{m}/0.1)(m/10^8)$ & s \\
$1/\nu_{\rm A}\equiv \tau_2$ & $1.7\times 10^5 \eta^{-1}(m/10^8)$& s\\
$D_3/c$ &  $1.8\times 10^9(\dot{m}/0.1)^{5/3}(m/10^8)^{4/3}$ & s\\
$W_{\rm max}$ & $1.8\times 10^{23}q(\Gamma/20)(\dot{m}/0.1)^{4/3}(m/10^8)^{2/3}$ & eV \\
$L_{\rm tot}$ & $1.2\times 10^{45} (\dot{m}/0.1)(m/10^8)$ & erg
	 s$^{-1}$\\
$L_{\rm A}$ & $1.2\times 10^{42} \eta(\dot{m}/0.1)(m/10^8)$ & erg s$^{-1}$\\
$L_{\rm \gamma}$ & $ 1.2\times 10^{41}(\eta\kappa/0.1) (\dot{m}/0.1)(m/10^8)$ & erg s$^{-1}$\\
$L_{\rm UHECR}$ & $1.2\times 10^{40}(\eta\kappa\zeta/0.01)
     (\dot{m}/0.1)(m/10^8)$ & erg s$^{-1}$\\
$L_{\rm UHECR}/L_{\rm tot}$ & $1.0\times 10^{-5}(\eta\kappa\zeta/0.01)$ & --\\
$L_{\rm UHECR}/L_{\rm \gamma}$ & $1.0\times 10^{-1}(\zeta/0.1)$ & --\\
\hline
\end{tabular}
\end{center}
$\zeta=L_{\rm J}/L_{\rm tot}$, $\eta=\nu_A Z_{\rm D}/{c_{\rm A}}_{\rm
 D}$,
$\kappa=E_{\rm CR}/E_{\rm A}$, and
$\zeta=\ln (W_{\rm max}/10^{20}{\rm eV})/\ln (W_{\rm max}/W_{\rm min})$.
\end{table}

\begin{figure}
\begin{center}
\rotatebox{0}{\includegraphics[angle=270,scale=0.4]{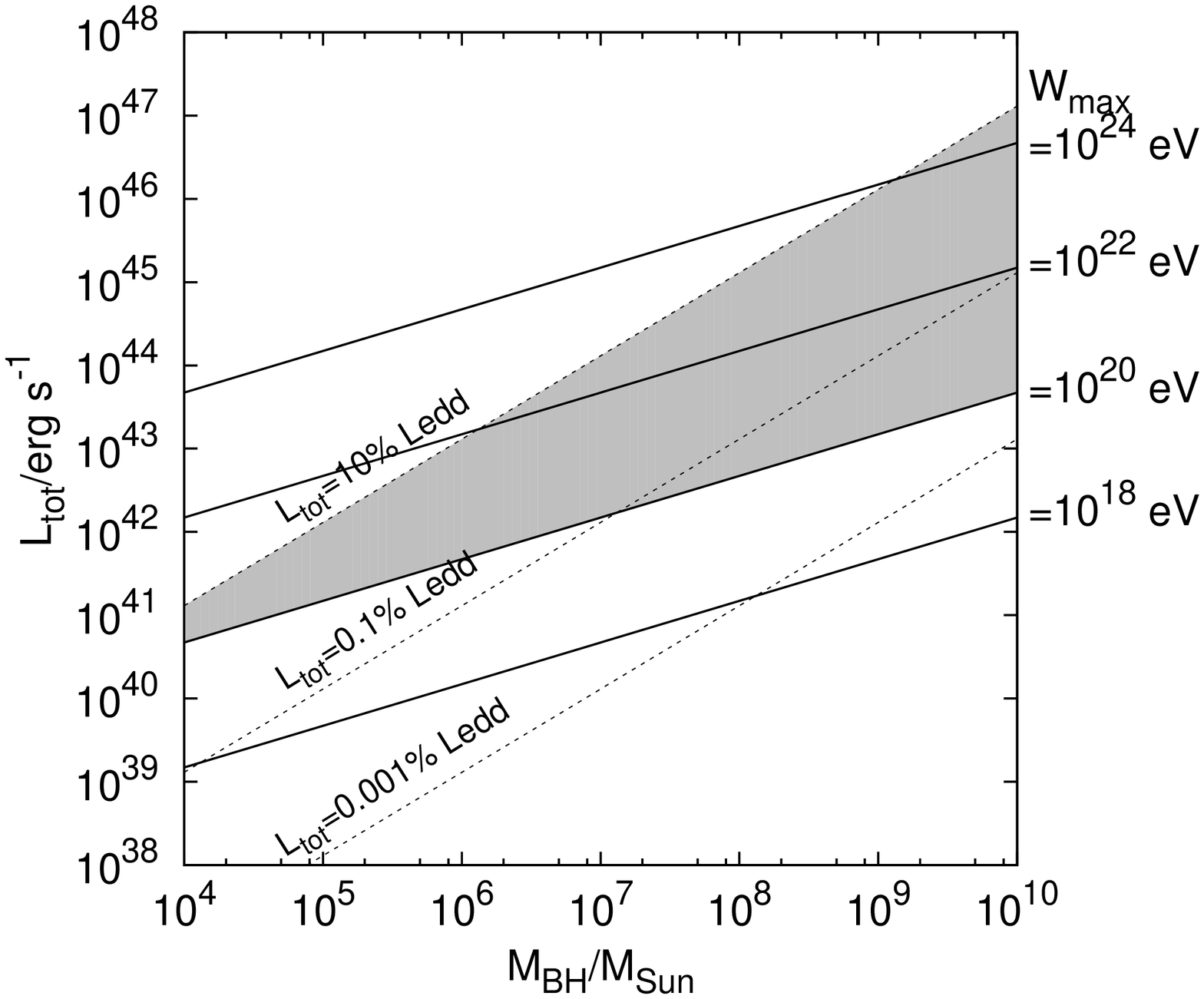}}
\caption{Revised version of Fig.~4 in \citet{Ebisuzaki14}
applying the radius $R=6.4{\rm R_g}=3.2{\rm R_s}$ instead of
$R=20{\rm R_g}=10{\rm R_s}$ as the magnetic field amplification site
for the standard disk model \citep{Shakura73}.
Solid lines represent maximum energy of accelerated particle
for the energies $W_{\rm max}=10^{18}, 10^{20}, 10^{22}$,
and $10^{24}$ eV
via ponderomotive acceleration
on the plane of
the central black hole mass and accretion power $L_{\rm tot}=\dot M c^2$,
assuming charge of the accelerated particle $q=1$,
and bulk Lorentz factor of the jet ($\Gamma=20$).
Dashed lines show 10\%, 0.1\% and 0.001\% accretion rate to the 
Eddington luminosity.
Since we do not consider any radiative processes,
i.e., RIAF ($L_{\rm tot} \le 10\%$ Eddington luminosity),
the gray shadowed region shows the objects
for the UHECR ($W_{\rm max}\ge 10^{20}$ eV) accelerators
by our models.
\label{MBH_Ltot}}
\end{center}
\end{figure}

\subsection{gamma-rays}
Blazar is a subclass of AGNs
for which the jets are close to our line of sight.
A blazar jet is very bright due to relativistic beaming effect,
including high energy gamma-ray bands.
They are observed in multi-wavelength from radio
to TeV gamma rays.
Short time variabilities and polarization
are observed, including recent {\it AGILE} and {\it Fermi} observations
for high energy gamma ray bands
\citep{Abdo09a,Abdo10a,Abdo10b,Abdo10c,Ackermann10,Striani10,Abdo11,Bonnoli11,
Ackermann12,Chen13,Ackermann16,Britto16}.

These non-thermal emissions are usually explained by
the internal shock model \citep{Rees78}, i.e.,
two shell collisions \citep{Spada01,Kino04,Mimica10,Peer16}
for which a rapid shell catches up a slow shell,
forming relativistic shocks.
At the shocks particle accelerations,
generating non-thermal particles, are expected
by Fermi acceleration mechanism.
Finally non-thermal emission is produced
via synchrotron emission and inverse Compton emission.
Other gamma ray emission from accretion
disks and related processes have been suggested
by \citet{Holcomb91,Haswell92}.

Our model can naturally explain properties of 
observed active gamma-ray flares, i.e.,
spectrum and timescales of flares.
For electrons energy loss via synchrotron radiation
causes a cutoff around PeV regime \citep{Ebisuzaki14},
although heavier particles, such as protons and heavier nuclei are
accelerated up to ultra high energy cosmic ray regime ($\sim 10^{20}$ eV)
and beyond.
The accelerated electrons emit radiation from radio to
high energy gamma-rays via synchrotron radiation and
inverse Compton emission mechanism.
The distribution of accelerated non-thermal particles
becomes power law with a power law index $\sim -2$ \citep{Mima91},
which is consistent with the observed blazar spectrum
with the power law index close to $-2$.
The photon index becomes close to $-2$,
when the light curve of gamma-rays becomes active phase
\citep{Abdo10a,Abdo11,Britto16}.
This anti-correlation between 
the gamma-ray light curves
and the photon index also supports our results.

From our numerical simulations
the rising timescale of electromagnetic flares
in the jet is as same as
rising timescale of magnetic field amplification in the disk,
i.e., typically $\bar \tau_1 \sim 30 GM_{\rm BH} c^{-3}$
and timescales of peak to peak in the flares
are as same as timescale of repeat cycle of magnetic field amplification
in the disk, i.e., typically $\bar \tau_2 \sim 100 GM_{\rm BH} c^{-3}$.

As for comparison with observed gamma-ray flares of blazars
the rising timescale of flares and
timescales of the cycle of the flares
are normalized by the $(1+z)GM_{\rm BH} c^{-3}$,
where $z$ is cosmological redshift of the object.
The redshifts for two objects are
$z_{\rm 3C454.3}=0.86$ \citep{Lynds67} and $z_{\rm AO0235+164}=0.94$.
From the observations of line widths of ${\rm H}\beta$
in the broad line region (BLR),
the mass of central black hole in 3C454.3
is estimated from $5\times 10^8M_{\odot}$ \citep{Bonnoli11}
to $4\times 10^9M_{\odot}$ \citep{Gu01}.
In this paper we adopt $5\times 10^8 M_{\odot}$
as a mass of central black hole in 3C454.3,
since the estimation of the BLR was done by using
C$_{\rm IV}$ line with less contamination by the non-thermal continuum
in \citep{Bonnoli11}.
The mass of central black hole in AO0235+164 is also derived to 
$M_{\rm BH(AO0235+164)}=5.85 \times 10^8 M_{\odot}$ \citep{Liu06}
from the line width of ${\rm H}\beta$ in the BLR.

We adopted 7 days as repeat time and
3 days as rise time for 3C454.3,
though various timescales with different times are observed 
from 3C454.3.
Among them, the seven days flare observed by \citet{Abdo11}
is the most energetic.
The estimated apparent isotropic gamma-ray 
energy in the seven days flare is 
4 times or more higher than those of flares reported
in \citet{Abdo10a,Britto16}.
Sub-energetic and shorter timescale flares reported
in \citet{Striani10,Ackermann12,Britto16} can be explained
the result of
the magnetic eruption via reconnection at the smaller region in the
accretion flows.

For AO0235+164 we compare the flare
reported in \citet{Abdo10a},
since the flare is the most energetic one
in the apparent isotropic energy
compared with other flares of AO0235+164,
for example \citet{Ackermann12}.
The rising timescale is three weeks
and repeat timescale is four weeks.
Table \ref{comparisontable} summarize of the comparison
between our results, theoretical model by \citet{Ebisuzaki14}
and observations.
Both timescales for  3C454.3  are good agreement with our results.
For AO0235+164 the timescales
are longer than those for our results
which suggests the magnetic field amplification may
occur outward where the timescale of MRI growth becomes longer.

%%%%%%%%%%%%%%%%%%%%%%%%%%%%%%%%%%%%%%%%%%%%%%%%%%%%%%%%%%%%%%%%%

\section{DISCUSSIONS and SUMMARY}
\label{sec:conclusions}
We have performed 3D GRMHD simulation of
accretion flows around a spinning black hole ($a=0.9$)
in order to study the AGN jets from
the system of supermassive black hole
and surrounding accretion disk
as an ultra high energy cosmic
ray accelerator via wake field acceleration mechanism.

We start our simulation from a hydrostatic disk,
i.e., Fishbone-Moncrief solution
with a weak magnetic field and random perturbation in thermal pressure
which violate the hydrostatic state and axis-symmetry.
We follow the time evolution of the system until $8300 GM_{\rm BH} c^{-3}$.
Initially imposed magnetic field is well amplified,
due to differential rotation of the disk.
Non axis-symmetric mode, i.e.,
bar mode near the disk edge grows up.
For highest resolution calculation case,
the typical timescale of the magnetic field growth near the disk edge
is $\bar\tau_1 = 30 GM_{\rm BH} c^{-3}$
which corresponds to inverse of the growth rate
of the MRI for the almost fastest growing mode.
For lower resolution calculation case,
the time scale of the magnetic field growth near the disk edge
becomes longer one.
And the thickness of the filamental structure in the magnetic pressure
around the equator increases by the lower resolution calculations.
Amplified magnetic field once drops then grows up again.
The typical repeat timescale is $\bar\tau_2 = 100 GM_{\rm BH} c^{-3}$
which corresponds to 
the analysis by high resolution local
shearing box simulations.
The transition between
low $\beta$ state and high $\beta$ state repeats.
This short time variability for the growth of poloidal magnetic
field neat the disk edge also can be seen
in the mass accretion rate at the event horizon
which means the mass accretion seen in our numerical simulation
triggered by angular momentum transfer 
by the growth of the magnetic fields.

We have two types outflows as shown in Fig.~\ref{plasmabeta_TEST600}.
One is the low density, magnetized, and collimated outflow,
i.e, jets.
The other is disk winds which are dense gas flow
and between the disk surface and jets.
In the jet short time variabilities of the electromagnetic luminosity
are observed.
The timescales are similar with those seen in the mass accretion rate
at the event horizon and
and poloidal Alfv\'en energy flux in the disk
near the ISCO,
i.e., typical rising timescales of flares
and typical repeat cycle 
are as same as the rising timescales of the magnetic field amplification
and repeat cycle of the magnetic field amplification,
$\bar \tau_1 =30{\rm GM_{BH}}c^{-3}$ and 
$\bar \tau_2 =100{\rm GM_{BH}}c^{-3}$, respectively.
Thus short pulsed relativistic Alfv\'en waves
are emitted from the accretion disk,
when a part of stored magnetic field energy is released.
Since the strength parameter of these waves extremely high 
as $\sim 10^{11}$ for the $10^8$ solar masses central black hole
and 10\% Eddington accretion rate accretion flows,
the wakefield acceleration proposed by \citet{Tajima79}
can be applied in the jet
after mode conversion from Alfv\'en waves to electromagnetic waves.
There are some advantages against
Fermi acceleration mechanism \citep{Fermi54}.
When we apply this mechanism to the cosmic ray acceleration,
the highest energy of cosmic ray reaches
$10^{22}$ eV which is enough high to explain
the ultra high energy cosmic rays.
We observe magnetic field amplification occurs
inner radius compared with the assumption by \citet{Ebisuzaki14},
i.e., $R=20{\rm R_g}$.
If we apply the model by \citet{Ebisuzaki14}
assuming that magnetic field amplification occurs
much inside the disk.
The two timescales are consistent with our numeral results.

Since both protons and electrons are accelerated via
ponderomotive force in the jet.
High energy gamma-ray emission are observed
if we see the jet almost on-axis, i.e., blazars.
The observed gamma-ray flare
timescales such as rising timescales
of flares and repeat cycle of flares
 for 3C454.3 by {\it Fermi} Gamma-ray Observatory \citep{Abdo10a}
are well explained by our bow wake acceleration model.
Telescope Array experiment reported 
that there is hotspot for the cosmic ray with the energy higher than 57EeV
\citep{Abbasi14}.
This observation supports that AGN jet is the origin of cosmic ray.

Lastly, the consequences from our present
work include the following implication
on the gravitational observation. 
Since non-axis-symmetric mode grows in the disk,
mass accretion onto the black hole
causes the emission of gravitational waves \citep{Kiuchi11}.
We estimate the levels of the signal of these gravitational waves,
assuming the black hole and accretion disk system.
The dimensionless amplitude of the gravitational wave
at the coalescing phase
can be estimated as \citet{Matsubayashi04}:
\begin{eqnarray}
h_{\rm coal}=5.45\times 10^{-21}
\left( \epsilon_{\rm GW} \over 0.01 \right)
\left( 4 {\rm Gpc} \over R \right)
\left(\mu \over \sqrt {2} \times 10^3 M_{\odot} \right),
\end{eqnarray}
where  $\epsilon_{\rm GW}$ is efficiency and we here assume as
$1\%$,
$\mu$ is reduced mass of the unit of solar mass
for the black hole
and the blob $\sim$ mass of the blob.
The mass of the blob is estimated
by $\dot M \bar \tau_1 $.
Figure~\ref{GW}~(a) shows time evolution of
estimated amplitude of the gravitational waves from our mass accretion rate
for the blazar 3C454.3.
The mass accretion history from $t=6500 GM_{\rm BH} c^{-3}$ to
$t=8300 GM_{\rm BH} c^{-3}$ is used for this plot.

\begin{figure}
\begin{center}
\rotatebox{0}{\includegraphics[angle=270,scale=0.28]{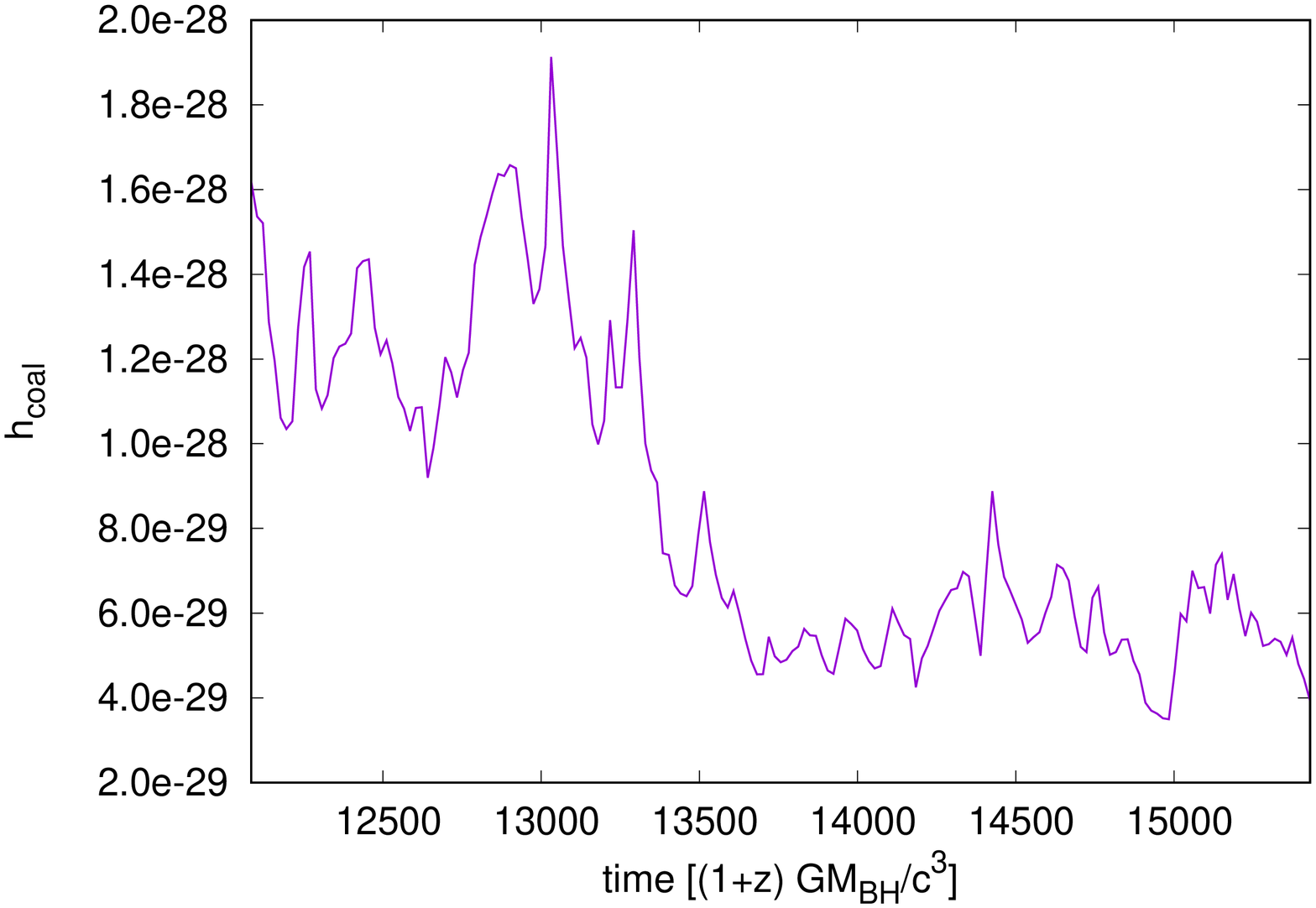}}
\rotatebox{0}{\includegraphics[angle=270,scale=0.28]{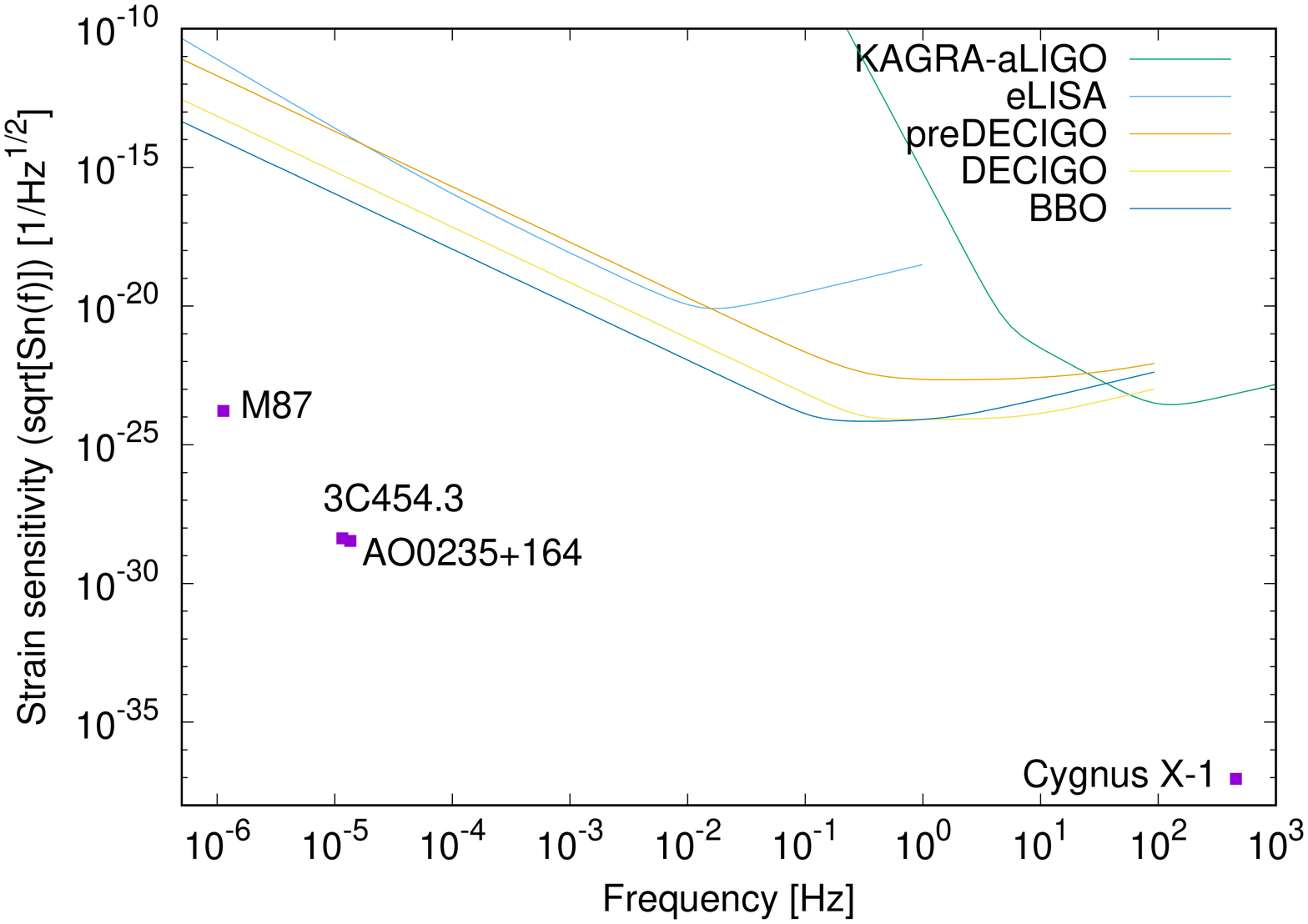}}
\caption{Top: Time evolution of amplitude of gravitational waves
for 3C454.3 ($z=0.86$) derived 
by the mass accretion rate shown in Fig.~\ref{magden}.
Bottom : Estimated gravitational wave signal when gas blobs
accrete onto the black holes.
Sensitivity curves
of grand based gravitational detector (KAGRA)
and proposed space gravitational detectors (eLISA, preDECIGO, DECIGO, and
 BBO)
are also presented.
\label{GW}}
\end{center}
\end{figure}

Figure~\ref{GW}~(b) shows the estimated amplitude of the gravitational
wave for some objects,
such as gamma-ray active blazars AO0235+164 ($=0.94$) and 3C454.3 ($z=0.86$),
nearby AGN jet M87, and famous stellar black hole Cygnus X-1.
We assume 
1\% Eddington mass accretion rate
and typical frequency of the gravitational wave
signal is $\bar \tau_1^{-1}/(1+z)$, where $z$ is redshift of the object.
Approximated sensitivity curves of the
KAGRA \citep{Nakano15}
proposed space gravitational wave detectors
such as eLISA \citep{Klein16},
preDECIGO \citet{Nakamura16},
DECIGO \citep{Kawamura06} and BBO \citep{Yagi11}
are also presented.
The signal level is so far small compared to the limit of the
presently operating or proposed gravitational antennas.

Our model cal be applied to 
magnetic accretion flows onto the central objects,
arising from other events
such as black hole or neutron star collisions.
For example,
recently the gravitational waves
from neutron star merger have been detected by LIGO and Virgo
gravitational wave detectors \citep{Abbott17a}.
Short gamma-ray burst followed this event just
1.7 s later of the two neutron stars merger \citep{Abbott17b}.
If the jets which emit gamma-rays
are powered by magnetic accreting flows
onto the merged object,
strong and relativistic pulses of Alfv\'en waves
would be emitted like our analysis of accretion disks
and then charged particles in the jet are accelerated
by the electromagnetic waves as \citet{Takahashi00} discussed.
We see the present acceleration mechanism and its
signature of gamma-ray bursts in an ubiquitous
range of phenomena.

\section*{Acknowledgements}
The authors wish to acknowledge the anonymous referee
for his/her detailed and helpful comments to the manuscript.
We thank Shinkai H. for introducing some references
for the fitting formula on the sensitivities of gravitational wave detectors.
We appreciate useful comments by K. Abazajian and B. Barish.
This work was carried out on Hokusai-greatwave system at RIKEN
and XC30 at CFCA at NAOJ.
This work was supported in part by 
the Grants-in-Aid of the
Ministry of Education, Science, Culture, and Sport 
15K17670 (AM), 26287056 (AM \& SN),
Grant-in-Aid for Scientific Research on Innovative Areas Grant
Number 26106006 (ET),
the Mitsubishi Foundation (SN), 
Associate Chief Scientist Program of RIKEN SN,
a RIKEN pioneering project `Interdisciplinary Theoretical Science
(iTHES)' (SN),
and 'Interdisciplinary Theoretical \& Mathematical Science Program (iTHEMS)' 
of RIKEN (SN).

%%%%%%%%%%%%%%%%%%%%%%%%%%%%%%%%%%%%%%%%%%%%%%%%%%



%\bibliographystyle{mnras}
%\bibliography{my} % if your bibtex file is called example.bib

%%%%%%%%%%%%%%%%%%%%%%%%%%%%%%%%%%%%%%%%%%%%%%%%%%

%%%%%%%%%%%%%%%%% APPENDICES %%%%%%%%%%%%%%%%%%%%%

%\appendix


%%%%%%%%%%%%%%%%%%%%%%%%%%%%%%%%%%%%%%%%%%%%%%%%%%

% Don't change these lines
\bsp	% typesetting comment
\label{lastpage}
\end{document}